\documentclass[11pt,titlepage]{article}

\usepackage{setspace} 

\usepackage{fullpage}

\usepackage{amsmath}

\usepackage{graphicx}

\usepackage[round,numbers,sort&compress]{natbib} 
\usepackage{subfigure}
 \usepackage{xcolor,graphicx}
 \usepackage{ulem}

\title{Spatial and Temporal Sensing Limits of Microtubule Polarization in Neuronal Growth Cones by Intracellular Gradients and Forces}

\author{Saurabh Mahajan,\\
	Div. of Biology, \\
	IISER Pune, Pune, India 
	\and Chaitanya A. Athale \thanks{
           Corresponding author.  Address: 
           Div. of Biology,
	   Indian Institute of Science Education and Research (IISER) Pune,
	   Sai Trinity, Sutarwadi Road, Pashan,
	   Pune, Maharashtra, 411021, India,
	   Tel.:~(+91-20)2590-8050, Fax:~(+91-20)2589-9790} \\
	Div. of Biology, \\
	IISER Pune, Pune, India}

\date{}

\begin{document}

\maketitle

\abstract{Neuronal growth cones are the most sensitive amongst eukaryotic cells in responding to directional chemical cues. Although a dynamic microtubule cytoskeleton has been shown to be essential for growth cone turning, the precise nature of coupling of the spatial cue with microtubule polarization is less understood. Here we present a computational model of microtubule polarization in a turning neuronal growth cone (GC). We explore the limits of directional cues in modifying the spatial polarization of microtubules by testing the role of microtubule dynamics, gradients of regulators and retrograde forces along filopodia. We analyze the steady state and transition behavior of microtubules on being presented with a directional stimulus. The model makes novel predictions about the minimal angular spread of the chemical signal at the growth cone  and the fastest polarization times. A regulatory reaction-diffusion network based on the cyclic phosphorylation-dephosphorylation of a regulator predicts that the receptor signal magnitude can generate the maximal polarization of microtubules and not feedback loops or amplifications in the network. Using both the phenomenological and network models we have demonstrated some of the  physical  limits within which the MT polarization system works in turning neuron.

}

\clearpage

\section*{Introduction}
%

A long-standing goal in developmental biology has been to understand the process of neuronal path finding that results in the complex wiring seen in the nervous system. Studies have unravelled multiple cues that provide directional signals to the axon. Once the directional cue is detected by a the growth cone (GC) at the end of an axon, the intracellular cytoskeleton undergoes polarized growth. The signaling networks that transduce an extracellular directional cue into cytoskeletal polarization during a neuronal GC guidance show striking similarities with those in highly migratory cell types such as slime molds {\it Dictyostelium}, neutrophils and migratory fibroblasts  \citep{parent1999}. This diversity of cell types suggests an evolutionarily conserved set of mechanisms in cell guidance. Experimental work in nerve GCs has shown both actin and microtubule dynamics to be essential \citep{mitchison1988}. While many studies have investigated the role of actin in filopodial and lamellipodial dynamics in neuronal growth cones, the precise role of MTs is less well studied (reviewed in \citep{gordonweeks2004}). MTs have been hypothesized to function like a cellular compass, which by random fluctuations searches directional space. In the presence of a cue, a reinforcement of the direction by transport of actin polymerization factors and membrane vesicles leads to turning of the GC \citep{andersen2005}. Thus MT polarization appears to be an important early event in GC turning, but consists of  complex interplay of chemical and mechanical effects.

The chemical regulation of MTs in neuronal growth cones is through modulation of MT dynamic instability- the property of MTs to transition from growing state to a shrinking state (catastrophe) and vice-a-versa (rescue) \citep{mitchison1984, hill1984}.  This dynamics has been shown experimentally to be essential for growth cone turning \citep{mitchison1988, tanaka1995a, tanaka1995b, challacombe1997, suter2004} and is most sensitive to the frequency of catastrophes \citep{4_param_model}. Important regulators of length dynamics is a family of neuron specific proteins of the stathmin (op18) family- stathmin 2 (SCG10) and stathmin 3 (SCLIP) \citep{poulain2007}. The protein SCG10 destabilizes microtubules by increasing catastrophe events and on phosphorylation results in stabilized MTs, corresponding to decreased catastrophe events \citep{ riederer1997, grenningloh2004}, in a manner similar to stathmin. SCG10 levels in primary cultured neurons appear to be critical for maintaining a balanced level of MT dynamics to enable neurite extension \citep{morii2006a} and {\it vivo} experiments indicate the N-methyl D-aspartic acid (NMDA) receptor (NMDAR) activation regulates SCG10 phosphorlyation through ERK2 \citep{morii2006b}.

Mechanical force generated by the retrograde flow of actin flowing inwards from filopodial tips regulates the number and velocity of MT ingress into filopodia in neuronal growth cones \citep{schaefer2002}. The coupling of the actomyosin driven flows with the MT system is thought to occur through discrete sites of binding by factors with either both actin- and microtubule-binding activity (e.g. formins) or complexes of factors that have such activity \citep{MT_Actin_Coupling_Molecules}. An additional force acts in the opposite direction of the retrograde flow driven by cytoplasmic dynein which works to both translocate microtubules towards the neuronal periphery, and to hold them in place in the filopodia \citep{Dynein_MyosinII}.

Theoretical models and simulations have been used to model some sub-cellular aspects of neurons providing insights into their dynamics. The cytoskeletal dynamics in growth cone polarization was modeled in filopodial dynamics were modeled to be randomly searching space \citep{vanveen1994a}. A model of microtubule assembly and tubulin transport could reconcile the constant elongation rates of neurites with tubulin assembly processes \citep{vanveen1994b}. Complex two-dimensional models which include the retrograde flow of actin suggest a non-random mode of MT invasion of filopodia \citep{hely1998}. A recent  detailed model of retrograde actin flow and myosin contractility has revealed the balance of forces driving the flow in neurons \citep{craig2012}. However all these models ignore the chemical regulatory elements that play a crucial role in neuronal cytoskeletal dynamics. Independently, chemical spatial gradients have been modeled in MT spindle assembly \citep{kalab2006} and cell polarization \citep{stathmin_grad}. Previous modeling of reaction-diffusion gradients and their regulation of MT dynamics has been used to explain the non-random growth of MTs in mitosis \citep{wollman2005, athale2008}. More recently, models of signaling proteins in neuritogenesis have demonstrated the role of feedbacks \citep{aoki2007} and of cell shape in axon  determination \citep{neves2008}. However a neuronal growth cone model that includes both regulatory reaction-diffusion components and the explicit MT dynamics and mechanics has not yet been developed. Given the accumulating evidence for the importance of microtubule regulation in growth cone guidance, a spatial model that includes both MT and actin interactions, combined with spatial dynamics of signaling proteins that modulate MT growth will be useful both from an experimental as well as theoretical perspective.

Here we develop a model based on {\it Aplysia} growth cone geometry and dynamics with experimentally measured parameters. Such a model provides on the one hand a platform for testing experimental hypotheses that can lead to extremely involved experiments. We develop two kinds of models: a phenomenological MT stabilization model as well as an explicit biochemically realistic reaction-diffusion model of MTs in turning GCs. Some of the typical scenarios of GC turning by a diffusible cue and perturbations of the actomyosin system show good agreement with existing experiments. Additionally we predict the minimal spatial and temporal detection limit. The maximal MT polarization angle shows good agreement with the maximal GC turning angle seen in experiments. The detailed biochemically realistic regulatory reaction-diffusion network model biologically is based on the cyclic phosphorylation-dephosphorylation of a microtubule destabilizer of the stathmin family, found in neurons. It predicts the most effective point of regulation in such a network. We have used both the phenomenological and network models to understand the  physical  limits within which the MT polarization system works in neuron that have detected a polarizing cue.

\section*{Theory}


\subsection*{Growth cone and axon shaft geometry}
The neuronal growth cone and terminal axon shaft were modelled as a 2D objects. The geometry was inferred image analysis of published images of {\it Aplysia} bag cell neuronal growth cones \citep{burnette2007} using ImageJ \citep{schneider2012} and a routine developed in-house using MATLAB R2008b (Mathworks Inc., USA). A circle of radius $25 \mu m$ is used to represent the growth cone based on dimensions of {\it Aplysia} growth cones. The growth cone is divided into concentric circles with the outer peripheral P-domain (radius 10 $\mu m$), the intermediate region transition T-zone (radius 5 $\mu m$) and the innermost central C-domain (radius 10 $\mu m$). The confining boundary is modeled as set of connected points with stiffness set to 100 $pN/\mu m$, making it effectively rigid. A rectangular region representing the axon-shaft emerges from the lower end of the circular growth cone. Based on reports of microtubule bundling by actomyosin forces in the growth cone neck \citep{bundling_forces} continuous spatial force field (Fig.~\ref{fig:dynfield}) was implemented. For every fiber model-point lying in this field, the value of the force at the corresponding position in the field is added to the net force on the point.

\subsection*{Microtubule orientation and dynamics}
Microtubules were implemented as elastic, non-extensible, polar fibers represented by a set of discrete model-points. The flexural rigidity (or bending stiffness) of microtubule fibers was set at 20 pN-$\mu m^2$ \cite{gittes1993, mickey1995}, which corresponds to a persistence length of 4.67 mm. Bending of these fibers at a shorter length scale is however possible based on the rigidity and force applied \cite{holy1997, laan2012}.  To maintain the simplicity of the model, we do not model microtubule breakage. The fibers were modelled to be dynamic based on the four-parameter model  \citep{4_param_model}: growth ($v_g$),  and shrinkage velocities ($v_s$), frequency of rescue ($f_{res}$) and catastrophe ($f_{cat}$). The parameters were spatially inhomogeneous with dynamic instability parameter values different for each of the domains as described in Table \ref{tab:mtdynpar}. These parameters affect only that plus-tip that enters it. An additional cortical catastrophe was modelled within 1$\mu m$ of the membrane periphery of the growth cone.\\

A fixed number of free microtubules ($N_{MTs}=30$) are initialized with lengths randomly assigned between 16-24 $\mu m$ by a uniform random number generator with a total angle of $106^o$ symmetrically about the vertical axis in an arc centered below the center of the circular growth cone (Fig.~\ref{fig:dynfield}). This geometry is based on {\it Aplysia} growth cone micrographs. The modeled MTs are dynamic at both ends. The parameters of dynamic instability: frequencies of catastrophe and rescue, and velocities of growth and shrinkage are distributed inhomogeneously as a function of radial position in the growth cone (Fig.~\ref{fig:dynfield}).  Inside the axon-shaft, minus ends of the modelled MTs undergo rapid catastrophe. The field of dynamic instability is set to a high $f_{cat}$ value (de-stabilizing) in the P-domain, while the T-zone and C-domain have a stabilizing effect (Fig.~\ref{fig:dynfield}). 

The minus ends of the MTs were initialized  20 $\mu m$ below the point of contact of the axon shaft with the growth cone in the axon shaft. The minus-ends are dynamic and undergo catastrophes if they move beyond $20 \mu m$ radial distance from the center due to a spatial change in {he catastrophe frequency to $f_{cat} = 1 s^{-1}$, while the other parameters remain constant (shrinkage velocity $v_{s} = 0.5 \mu m/sec$, rescue frequency $f_{res}= 1s^{-1}$ and $v_g=0 \mu m/s$).

\subsection*{Force due to actin retrograde flow in the P-domain}
The actomyosin retrograde flow is modeled by discrete immobilized motors arranged radially along filopodial bundles that push MTs inwards towards the C-domain (Fig.~\ref{fig:mt-actin}) based on reports in literature of coupling-molecules that can bind to MTs and actin simultaneously. In filopodia the actin itself is flowing inwards in a myosin-dependent retrograde flow, hence generating many point forces that pull MTs into the center of the growth cone. 
Recent reports on the role of kinesins in MT growth and polarization in the P-domain suggest that both the actomyosin system, and a kinesin based retrograde force are involved in moving MTs inwards from the periphery \cite{Kinesin_GC_turning}. Assuming both these systems act along the same direction, the force vectors due to actomyosin and kinesin activity can be added up and resulting in a high stall force value ($f_{stall}$) (Table  \ref{tab:motorparam}). Control calculations showed this force is sufficient to move MTs of a fixed length inwards with a retrograde flow speed that agrees with experimental measurements \citep{schaefer2002}.

 Motors-MT interactions are modeled as a Hookean springs where force exerted by a motor on a point of the fiber is given by $f_{ex} = k*\delta r$, where $k$ is the stiffness and $\delta r$ is the distance separating the motor and the point on the fiber. Motors were characterized by their positions, attachment/detachment rates ($r_{attach}/r_{detach}$), speed of movement ($v_{mot}$), direction of movement, stall force ($f_{stall}$) and stiffness ($k$) of the induced links. Detachment rate depends on the force generated as modelled previously  \citep{klumpp2005} as $r_{detach}=r_{detach}^{'}*exp^{(2*\frac{|f_{ex}|}{f_{stall}})}$ based on Kramers theory \citep{kramers40}. The filopodial arrays (n= 61) are arranged between 0 and $180^o$ with respect to the positive X-axis through the semi-circle of the GC with 30,000 motor complexes randomly distributed (uniformly) amongst the filopodial bundles $\sim 500$ motors per bundle). The speed of these motor complexes was set to 0.07 $\mu m/s$ while attachment-detachment rates were optimized (see Table \ref{tab:motorparam}) to provide net translocations velocities in agreement with measurements in {\it Aplysia} growth cones \citep{schaefer2002}. 
 
In the T-zone an analog of immobilized non-motile dynein motors is modelled with $r_{attach} = 5$, $r_{detach} = 0.1$ and $v_{mot} = 0$. These motors hold MTs but do not move the microtubules when bound (Fig.~\ref{fig:mt-actin}).

\subsection*{Reaction-diffusion network model of receptor driven intracellular polarization}
A simple reaction-diffusion network consisting of three species: two cytosolic diffusive species assumed to be stathmin $S$ and its phosphorylated form $S^*$ and a membrane immobilized receptor species (R). An optional positive-feedback interaction by growing microtubule plus tips ($N_{tips}$) and a weight of the feedback ($w_{1}$) was implemented. This feedback increases the $S->S^*$ conversion in presence of growing plus-tips. $S$ is hypothesized to be phosphorylated to $S^{*}$, similar to stathmin phosphorylation, near the membrane by the receptor $R$, mimicking the NMDA-Erk2 pathway. The backward reaction converts $S^*$ to $S$ homogeneously in the cytosol governed by a first order rate constant $k_2$. In the positive-feedback model, growing MT plus tips can also enhance this forward reaction. The forward reaction is modelled using Michaelis-Menten kinetics as follows:

\begin{equation}
\frac{d[S]}{dt} =  - \frac{k_{1}[R][S]^{n_1}}{K_{M1} + [S]^{n_1}} - N_{tips}\cdot \frac{k_{fb} \cdot w_{1}\cdot [S]^{n_{2}}}{K_{M2} + [S]^{n_{2}}} + k_{2}[S^{*}] + D_{S}\cdot\nabla^{2} [S] 
\label{eq:s}
\end{equation}

The backward reaction is modelled using first-order kinetics as:
\begin{equation}
\frac{d[S^*]}{dt} =    \frac{k_{1}[R][S]^{n_1}}{K_{M1} + [S]^{n_1}} + N_{tips}\cdot \frac{k_{fb}\cdot w_{1}\cdot [S]^{n_{2}}}{K_{M2} + [S]^{n_{2}}} - k_{2}[S^*] + D_{S^*}\cdot\nabla^{2} [S^{*}] 
\label{eq:s*}
\end{equation}\\

The terms $n_1$ and $n_2$ correspond to cooperative Hill-coefficients for the forward reaction and the MT-dependent feedback terms respectively. These terms represent cooperativity in phosphorylation ($n_1$) or the positive feedback ($n_2$), due to either the extracellular regulated kinase (ERK) pathway \cite{levchenko2000} or multiple sites of phosphorylation in stathmin \cite{manna2009}. Unless explicitly stated, the Hill-coefficient parameters were set to $n_1=n_2=1$. Diffusion coefficients of the two diffusing cytosolic species $S$ and $S^*$ are given by $D_S$ and $D_{S^*}$ while the concentration gradient is indicated by the 2D Laplacian $\nabla^{2}$.

The role of S in modulating MT dynamics was implemented by increasing the local catastrophe frequency $f_{cat}^{'}$ dependent on [S] using a phenomenological expression described previously based on {\it in vitro} data \citep{stathmin_cat}:
\begin{equation}
f_{cat}^{'} = f_{cat}\cdot exp(w_{2}\cdot([S] -[S_{tot}]))
\label{eq:fcat}
\end{equation}
Here $f_{cat}$ is the default value (Table \ref{tab:mtdynpar}), when $[S]=[S]_{tot}$, $[S]_{tot}$ is the sum concentration of modified and unmodified stathmin, and $w_2=0.3144$, a scaling factor obtained by fitting to experimental data \citep{stathmin_cat}. The parameters used for the reaction-diffusion network are taken from literature and described in Table \ref{tab:rd-parameters}.

\section*{Methods}

\subsection*{Simulation of the microtubule model}
Our model is implemented and simulations conducted using Cytosim- a cytoskeleton simulation engine implemented in C++ \citep{cytosim}. All simulations were run for 1200 seconds (or more). Typical simulations include 30 dynamic MTs, 30000 surface-immobilized motors in P-domain and 1000 grafted motors in T-zone. The reaction diffusion system was simulated with spatial resolution of $1 \mu m$. Positions of all objects and occurrence of events like movement, binding/unbinding, change in length of MTs (due to growth/shrinkage), change in MT end state (due to catastrophe/rescue) is calculated for all objects at each time step from the corresponding rate parameters. Simulations were run on a 12-core Dell T5500 workstation with 2.40 GHz processors using a publicly available parallel processing code (Distributed Parallel Processing Shell Script vers 2.85). Each typical run takes $\sim1500$ seconds to complete.

For each iteration total MT numbers, mean direction of MT distribution and circular spread of MTs were calculated in the P-domain and angular trends were analyzed in 12 angular sectors, $15^{o}$ each (Fig.~\ref{fig:gcdyn3}).  The mean angle of MT distribution in a growth cone at a time point (t) was calculated as:

\begin{equation}
\theta_{av}(t)=\frac{\Sigma{N_{tips}^{i}\cdot\theta_i}}{\Sigma{N_{tips}^{i}}}
\label{eq:theta_av}
\end{equation}

where $\theta_i$ is the mid-angle and $N_{tips}^i$ are the number of plus-tips in the $i_{th}$ sector. The corresponding circular standard deviation ($\sigma_c(t)$) is given by 

\begin{equation}
\sigma_{c}(t)= \sqrt{ \frac{\Sigma{N_{tips}^i} \cdot (\theta_{i} -{\theta}(t))^2}{\Sigma{N_{tips}^i}} }
\label{eq:csd}
\end{equation}

Time averages of mean angle ($\langle\theta_{av} \rangle$) and circular standard deviation ($\langle\sigma_{c}\rangle$) were used as measures to compare with experiments. In experiments where a directional bias was introduced, the difference between mean angles averaged at steady state before and after the bias ($\Delta\langle\theta_{av} \rangle$) were evaluated as a measure of MT polarization.

\subsection*{Simulation of the reaction-diffusion model}
The 2D system of PDEs was solved on a finite lattice grid using by an explicit forward difference scheme \citep{num_math_book}. The spatial interval $\Delta x=1 \mu m$ and time interval $\Delta t=0.01 sec$ were optimized by comparing simulation results with analytical solutions of known initial condition cases (Fig.~\ref{sup:rd-optimization}) \cite{crank1976}. Zero-flux boundary conditions were used at the growth cone and axon-shaft boundaries (Fig.~\ref{fig:dynfield}). All preliminary calculations were performed in Octave \citep{eaton2002} and then implemented in C++.

\subsection*{Analysis of experimental MT distributions}
Images of fluorescently labelled microtubules in {\it Aplysia} growth cones were analyzed in house using a MATLAB script (Mathworks Inc., USA) and ImageJ (NIH, USA) combined with manual counting of plus tips.

\section*{Results}


\subsection*{Microtubule angular distribution is uniform and fluctuating}
A two-dimensional model of the {\it Aplysia} growth cone with dynamic MTs has been developed to understand the quantitative details  of spatial remodeling of MTs in the earliest events relating to neuronal growth cone turning. The model integrates MT regulatory and mechanical components as described in the model section. The simulation geometry with spatial dynamic instability and filopodial geometry is shown in Fig.~\ref{fig:dynfield}. A radial inward-directed retrograde flow and dynein based stationary coupling of MTs is modeled using a motor model geometry schematically illustrated in Fig.~\ref{fig:mt-actin}. A simulation run of such a model shows MTs entering the P-domain in $\approx10$ min to attain a uniform but fluctuating distribution (Fig. \ref{fig:gcdyn1}, Movie~\ref{sv1}). We also observe some MT buckling mostly in the T-zone. This geometry and dynamics are qualitatively similar to {\it in vitro} data of {\it Aplysia} growth cones. The quantitative dynamics of MTs entering the P-domain also show rapid steady state where on an average $2/3^{rd}$ of the MTs remain in the P-domain (Fig.~\ref{fig:gcdyn2}). The fluctuations suggest that MTs can constantly search the P-domain for directional cues, primed for turning. The number of MTs in P-domain for two different values of initialized MT numbers (30 and 60) show between 25-50 MTs on an average (Fig.~\ref{fig:gcdyn3}). This is in agreement with previous experimental observations \citep{schaefer2002, bundling_forces}. The spatial patterns of these MTs were analyzed by sampling the number of plus-tips in discrete angular periods for each time point (Fig.~\ref{fig:gcdyn4}). The weighted mean angle for the MTs in the P-domain of a GC was calculated based on Equation \ref{eq:theta_av} and plotted against time (Fig.~\ref{fig:gcdyn5}). After initial fluctuations reaches $\theta_{av}$ reaches a steady state value of $\approx 90^o$ in about 300 s of simulations, while the corresponding  circular standard deviation ($\sigma_c \approx \pi/6$) calculated from Equation \ref{eq:csd} achieves a steady state value of $\sim 45^o$. The time-averaged mean MT angle ($\langle\theta_{av}\rangle$) is as expected $\sim 90^o$ ($\langle\sigma_c\rangle \sim 30^o$),  independent of the number of MTs simulated (Fig.~\ref{fig:gcdyn6}). Both the mean angle and the circular standard deviation are in good agreement with experimental measures \cite{schaefer2002, bundling_forces}. Simulations of MT density dynamics in four equal sectors of the growth cone for initial MTs N=30  (Fig. \ref{fig:nmt30}), N=60 (Fig. \ref{fig:nmt60}) and N=90 (Fig. \ref{fig:nmt90}) are all of the same order of magnitude as experiments with {\it Aplysia} growth cones \cite{PKC_effect}.

In order to test if the standard model of motor detachment based on Kramers theory affects our results, a simple model with constant rates of motor detachment in the P-domain was tested. The values $r_{detach}= 1 s^{-1}$ or $r_{detach}= 5 s^{-1}$ are the two extreme values from a two-state load dependent detachment model which was shown to better fit experimental data for kinesin \cite{driver2011}. The mean MT number in the P-domain and the mean angle remain unchanged (Fig. \ref{sup:detachmodel}), suggesting little effect of changing detachment rates.

Thus our model simulations show qualitatively and quantitatively expected behavior in agreement with the available spatial and dynamic data. In the following sections we perturb this model system and compare it with experiments.

\subsection*{A decreased retrograde flow rate results in increased MT flux in the P-domain}
The role of the actomyosin retrograde flow was explored in our modeled growth cones by analyzing its effect on the flux of MTs entering the P-domain. A reduction in the retrograde flow rate (from $4.2 \mu m/min$ to $2.1 \mu m/min$) by $50\%$ results in a two-fold increase in MT-tip flux only in the peripheral $25\%$ of the modeled growth cone (Fig.~\ref{fig:flux-retrograde1}). Such a dramatic effect is not observed if larger proportions ($50\%$ and $75\%$) of the P-domain were sampled. This suggests the fluctuations in the most distal region of the growth cone are most sensitive to such treatment. On the other hand a two-fold increase in retrograde flow rate results in an $\approx 85\%$ decrease in flux of MT tips in the distal $25\%$ of the P-domain. This effect is also seen in experimental analyses of MT dynamics which report MT fluxes only in the distal $25\%$ of the growth cone. The effect of changing retrograde flow rates on absolute number of MTs in the P-domain shows small but statistically significant differences (Fig.~\ref{fig:flux-retrograde2}). The relative proportion of simulated MTs (N=30) compared to those entering the P-domain also progressively diminishes if we observe the most peripheral $25\%$ of the growth cone. Experimental measurements of the flux of MT plus-tips in with neuronal growth cones perturbed by blebbistatin - a myosin II ATPase inhibitor- indeed showed a $50\%$ decrease in retrograde flow rates results in a two-fold increase in the flux of MTs entering the P-domain in the peripheral $25\%$ of the P-domain \citep{burnette2007}. This correspondence with experimental data shows our model of retrograde flow can be used for further studies on MT dynamics modulation.

\subsection*{A spatial catastrophe frequency change is most effective in changing the polarization angle of MTs}
Recent evidence has been accumulating for the role of MT polymerization modulation in growth cone turning. We have systematically tested the role of MT dynamics in MT polarization, an early even in growth cone turning. Values of dynamic instability were modified in only one half of the P-domain of the simulated growth cone. The result of this spatial bias was evaluated by plotting a time course of the $\theta_{av}(t)$.  Although fluctuations in the mean MT-angle still remain large, a small change in mean angle is observed from the unbiased base (Fig.~\ref{fig:polarization-theta1}). Both the actual simulation (Movie ~\ref{sv2}) and a time averaged angular distribution of the MTs before and after bias in the control case compared to the biased case represent this change more graphically (Fig.~\ref{fig:polarization-theta2}). In order to test the parameter to which the system was most sensitive, the bias was systematically introduced in each of the parameters of dynamic instability ($f_{cat}$ and $f_{res}$). The change in mean angle ($\Delta \langle \theta_{av}\rangle$) of MTs in the outer half of the P-domain was evaluated and compared to the control (cat/res) (Fig.~\ref{fig:polarization-theta3}). Increasing catastrophe frequency turns the mean angle of MT distribution away from the direction of bias, while the increase in rescue frequency turns the MT distribution towards the bias. The ten-fold increase in $f_{res}$, ten-fold decrease in $f_{cat}$ and ten-fold increase in $f_{cat}$ show the significant changes in mean angles compared to the unbiased case (Two-way Student's t-test $p<0.05$). A ten-fold decrease in $f_{res}$ does not result in a significant change in mean angle of MTs. To further examine the minimal range within which the rescue and catastrophe frequencies change mean angles of MT, step-wise increases in  $f_{res}$ and $f_{cat}$ were made. Statistically significant changes in $\Delta\langle\theta_{av}\rangle$ were observed even with two- and four-fold changes in $f_{cat}$ and $f_{res}$ values (Fig.~\ref{fig:polarization-theta4}). These results suggest even a small change in dynamic instability can bias a the MTs in the given geometry, however changes in the $f_{cat}$ values produce the most dramatic effect on $\Delta\langle\theta_{av}\rangle$. These results are consistent with previous non-spatial models of MT dynamics that demonstrated theoretically as well as experimentally (in mitosis) that the mean MT length is most sensitive to changes in $f_{cat}$ \citep{4_param_model}. This also sets the stage for testing the spatial limits of MT polarization in our model growth cones.

\subsection*{Spatial and temporal limits to MT polarization sensitivity}
In order to test the sensitivity of the MT polarization system in the modelled growth we modeled different extents of the polarizing cue and evaluated the output.The sensitivity was estimated in two ways- (i) the change in mean angle ($\Delta \langle \theta_{av}\rangle$) as a function of the angular extent of the bias, and (ii) the time taken for the adaptation, i.e. the new steady state. In simulation $f_{cat}$ values were decreased or $f_{res}$ values were increased ten-fold in sectors of the P-domain. These sectors were all centered around $120^{o}$ and the angular extent was varied between $0-60^{o}$. The change in mean angle $\Delta \langle \theta_{av} \rangle$ was reduce from $\approx15^{o}$  to $\approx 5^{o}$ in response to both changes in catastrophe and rescue frequencies. The change in mean angle with decreasing spread of $f_{res}$ showed a seemingly steady polarization angle of $>10^o$ for bias extents even as small as $30^o$. A rescue frequency bias less than $30^o$ also resulted in a small but statistically significant changes (Student's t-test) in the mean angle (Fig.~\ref{fig:angleExtent2}). On the other hand reducing the spatial extent of $f_{cat}$ resulted in a steady reduction in $\Delta \langle \theta_{av}\rangle$ with ever decreasing spatial extent (Fig.~\ref{fig:angleExtent1}). This result appears to contradict the observed maximal sensitivity of MT length to the catastrophe frequency. It is likely to be a result of the increased catastrophe frequencies at the GC periphery, which are better compensated by a higher rescue frequency. This prediction can also be experimentally tested, and would suggest alternative sensitivity of the MT system in realistic cell geometries as compared to {\it in vitro} or mitotic measurements. Additionally it suggests that a measurable MT polarization cannot occur if the bias is smaller than $20^{o}$ in angular extent for catastrophe- and $10^{o}$ for rescue-frequencies. 

In order to test the temporal sensitivity of the polarization of MTs, one half of a growth cone was biased by changing either $_{cat}$ or $_{res}$ abruptly. The time taken for adaptation was estimated by fitting the dynamics of the mean angle of MTs in the growth cone ($\theta_{av}(t)$) to a simple saturation kinetic expression (Fig.~\ref{fig:angleExtent3}):

\begin{equation}
\theta_{av}(t) = \theta_{av}(0)+ \Delta \theta_{av}(t)\cdot \frac{t^n}{t^n + K_{t}^n}
\label{eq:theta_hill}
\end{equation}\\

where $\theta_{av}(t)$ is the mean direction at time-step $t$, $\theta_{av}(0)$ the mean direction before the bias is introduced, $\Delta\langle \theta_{av}\rangle$ is the net change in direction after the bias, $n$ is the Hill coefficient, and $K_t$ is the half-time within which polarization is half complete. The change in $K_t$ from the time of imposition of the bias is $\Delta K_t$. A smaller value of $\Delta K_t$ indicates a faster response to the stimulus. Ten-fold increases and decreases (independently) in catastrophe and rescue frequencies produced $\Delta K_t$ values ranging from 30-45 s. A ten-fold decrease in rescue frequency could not be fitted by the routine. This appears to indicate a characteristic time for half-growth cone bias experiments for MT polarization to be in the range of $\approx 30 s$. This could be experimentally tested in a carefully designed experiment.
At the same time, the phenomenological gradient of stabilization of MT growth in space must have an underlying mechanistic basis. Hence we developed a spatially explicit reaction-diffusion model of a potential MT regulatory protein in the framework of the previously developed spatial MT dynamics in the neuronal growth cone.

\subsection*{A cyclical phosphorylation-dephosphorylation reaction-diffusion network leads to MT polarization}
In order to test the role of a biologically realistic factors that can change microtubule dynamics, a reaction-diffusion spatial model of an MT regulator was developed, based on the role of the protein SCG10, the stathmin (op 18) homolog. We have simplified the network of extracellular regulatory kinase (ERK) dependent phosphorylation of stathmin to consider its core element, namely MT spatial regulation. In our model stathmin ($S$) phosphorlyation ($S^*$) occurs at the membrane due to a receptor ($R$) while dephosphorylation of can occur throughout the growth cone, resulting in a gradient of $f_{cat}$ (Fig.~\ref{fig:reac-scheme1}). An additional feedback term ($w_{1}$) regulates further amplifies the phosphorylation reaction. Both $S$ and $S^*$ are modeled to be diffusible through the growth cone geometry, while $R$ is a parameter in the model, as seen in Equations \ref{eq:s} and \ref{eq:s*}. In order to test numerical stability of our integration scheme, the diffusion of a single species was compared to that in an analytical model (Fig.~\ref{sup:rd-optimization}). The modulation of MT dynamics occurs in our model by modifying the $f_{cat}$ values based on Equation \ref{eq:fcat}. The resultant steady state distribution in such a simple model of phosphorylation-dephosphorlylation cycles demonstrates the differences in spatial extent of $S$ and $S^*$ in response to a stimulus in one half of the growth cone as seen along the radial (Fig.~\ref{fig:reac-scheme2}) and central (Fig.~\ref{fig:reac-scheme3}) axes. The time course of gradient formation by depletion of S in one half of the GC is visible in Movie ~\ref{sv3}.

In order to test which component of this reaction scheme might affect the polarization of MTs maximally, we compared the $\Delta \langle\theta_{av}\rangle$ for receptor signal strength and nature of the reaction kinetics. We tested the effect of changing the phosphorylation reaction of $S$ to $S^*$ transition from a simple first-order ($K_{M1}/[S_{tot}]=2$), intermediate ($K_{M1}/[S_{tot}]=1$) or zero-order saturated ($K_{M1}/[S_{tot}]=0.1$) scenario. The effect of changing receptor concentrations [R] (1, 10, 100, 200 $\mu M$) was simultaneously tested for each of the $K_{M1}/[S_{tot}]$ values. We find a high receptor signal values lead to most significant changes in $\Delta \langle \theta_{av} \rangle$ (Fig.~\ref{fig:Rsig-var1}). This is also borne out by the mean steady state post-bias angular distributions of MTs (Fig.~\ref{fig:Rsig-var2}). All reactions were carried out with the Hill-Coefficient in the phosphorylation reaction set to $n_1 = 1$ (Equation \ref{eq:s}). For the situation of a high receptor stimulus ($[R]=100 \mu M$), when $n_1$ was set to higher values (2, 4 and 10) the polarization angle $\Delta \langle\theta_{av}\rangle$ did not change appreciably (Fig. \ref{sup:hillcoef}), indicating a negligible effect of the cooperative term.

The feedback from MTs to promote the phosphorylation reaction was simulated and compared to the model without feedback, in terms of the final readout for the $f_{cat}$ distributions. The simple model demonstrates a slight but clear directional bias, while the feedback model in contrast shows local inhomogeneities in $f_{cat}$, corresponding to the presence of growing plus-tips (indicated by arrows)  (Fig.~\ref{fig:feedback1}). As a result the mean post-bias MT distribution is visibly polarized in the simple reaction case as compared to the strong feedback model  (with both positive-feedback $w_1=200$ and Hill-coefficient $n_1=10$) (Fig.~\ref{fig:feedback2}).

Taken together, these results suggest a model of a stathmin-like phosphorlyation-dephosphorylation mechanism without feedbacks can indeed polarize microtubules in a neuronal growth cone geometry. It remains to be seen if such a gradient and downstream MT effect can be measured experimentally.

\section*{Discussion}
%
%
%
%

Cytoskeletal dynamics in GC turning has been increasingly seen as critical for understanding neuronal GC guidance \cite{dent2003} since multiple signaling pathways converge on the actin-microtubule system \cite{dent2011}.  However, very few of the details of this coupling between the signaling mechanisms with the cytoskeleton in neuronal growth cones are known. Previous studies have focussed either only on the mechanics of cytoskeleton or the dynamics of the signaling system. Our study for the first time explicitly considers this coupling between microtubule mechanics and signaling gradients in a neuronal growth cone. We focus on the role of the spatial regulation of MTs in a turning GC. Another feature of our model is its comparison with previously published data of MT polarization in an {\it Aplysia} neuronal growth cone, thus making many aspects of our model biologically relevant and testable. All input parameters of microtubule dynamics and actin retrograde flows are taken from neuronal studies, except in the case of microtubule-motor mechanics and stathmin, where the parameters are derived from {\it in vitro} studies. 

Our neuronal MT-polarization model is based upon the following assumptions: (1) a static extracellular guidance cue, (2) single step phosphorylation of the MT regulatory protein, ignoring intermediate steps  (3) a phenomenological non-linear mapping of stathmin concentration and microtubule dynamics and (4) the two-dimensional geometry of the growth cone with the mechanics of retrograde pushing of MTs. During the process of neuronal GC turning the assumption of a static guidance cue holds true since the time scale of remodeling of the MTs is an order of magnitude slower than receptor polarization. For instance MT polarization in {\it Aplysia} growth cones occurs within $\sim 10$ min \cite{PKC_effect, MT_Actin_Coupling_Molecules}, while receptor polarization of NMDAR takes only seconds \cite{wang2011, guirland2004}. We use this difference in time-scales to focus on the events slower than membrane-receptor polarization, i.e. downstream signaling and cytoskeletal regulation. The downstream signaling from these receptor signals either through GTPases or through kinase-phosphatase systems have a conserved modular structure, permitting a simplification of the cascade as a simple two-component cyclical reaction \cite{brown1999, kholodenko2000}. Using such a relatively simple single-step reaction cycle, we can also test multiple modes of amplification and feedback in a manner that is tractable. Such amplification steps are thought to be part of the mitogen activated kinase (MAPK) extracellular-regulated kinase (ERK) pathway\cite{levchenko2000}, upstream of the cyclical reaction. Although the Hill-cooperativity modeled in Equations \ref{eq:s} and \ref{eq:s*} has not been measured in neuronal GCs, this model allows us to explore the multiple levels at which potential amplification of the reaction network can occur. The modeled two-dimensional geometry was assumed to be a reasonable approximation of the growth cone based on micrographs of {\it Aplysia} bag neuron growth cones which are typically 20-30 $\mu m$ in diameter but only 2-3 $\mu m$ in height in the P-domain \cite{PKC_effect}. 

Using this model we compare our simulations to previously published experimental data. It is clearly seen that the mean angle and spread of the modeled neuronal MTs  are strikingly similar to the experimentally measured values from {\it Aplysia} growth cones in the absence of a directional cue \cite{schaefer2002, bundling_forces}.  The simulated quadrant-wise time dynamic of MT densities in the peripheral 25$\%$ were plotted for GCs with the initial number of MTs as 30 (Fig. \ref{fig:nmt30}), 60 (Fig. \ref{fig:nmt60}) or 90 (Fig. \ref{fig:nmt90}). Agreement with experimental measures of MT density in the peripheral P-domain \cite{PKC_effect} and quadrant-wise  dynamics \cite{suter2004} is seen only for the MT numbers 60 and 90. However the effective MT-flux rates in simulation (Fig. \ref{fig:flux-retrograde1}, Table \ref{tab:mtflux}) are similar compared to experimental values \cite{burnette2007}.  The comparison of the change seen in MT flux in the peripheral 25$\%$ of the P-domain after a 50$\%$ reduction in retrograde flow rates is further evidence that the model of microtubule dynamics and geometry appears to be valid, given the available data. The converse experiment of the increase in flux rates has been performed by siRNA inhibition of dynein in the GC which oppose the retrograde flow. In experiments this results in an $\sim 8$ fold reduction in MT's in the peripheral P-domain \cite{Dynein_MyosinII}, a difference not seen in our model. However the interpretation of the experiment as well as  differences between rat superior cervical ganglia neurons from {\it Aplysia} neuron sizes, could partly cause the difference. We also find a spatial bias in the retrograde flows velocities can  influence MT polarization in either two-fold increase or decrease in retrograde velocities (Fig. \ref{sup:retrovel}), making our results comparable to selective inhibition of kinesin in only one half of the GC which resulted in growth cone turning \cite{Kinesin_GC_turning}.

A surprising property to emerge from our model simulations was the buckling of microtubules in the T-zone (Fig. \ref{fig:gcdyn1}). It results from the combination of the growth cone geometry, immobilized motor complexes of the T-zone, and actin retrograde flow in the P-domain. Such MT buckling has been previously observed in experimental measurements \cite{schaefer2002}, and is thought to also result in breakage of MT fibers, which then either polymerize or are translocated towards the P-domain. In previous iterations of the simulation, it was found that the T-zone localization of immobile motor complexes was critical to maintain the orientation and spread of MTs as they enter the P-domain. Hence our model would appear to suggest a functional role for such a set of complexes, that at first sight appear passive. Changing these complexes into minus-end directed active motors increases the buckling of the MTs. A further quantification of this buckling both in experiment and theory might prove useful. These molecular motor assemblies might also play a role in the proposed polarity sorting of short microtubules in neurons proposed in qualiitative models before by Baas et al. \cite{baas2012, baas1996}. Our model could be used in future to test the specifics of such processes.

The sensitivity of the MT polarization system in response to changes in dynamic instability parameters was measured to give an idea of the ideal points of regulation. We found the system to be most sensitive to the frequency of catastrophe. However statistically significant changes in mean MT angle were also observed as a result of fold-changes in both catastrophe and rescue frequencies. The change in mean MT angle is observed to be $20-30^o$ for rescue and catastrophe frequencies. The experimental mean turning angle observed for {\it Xenopus} embryonic \cite{MT_stabilization_causes_GC_turns} and spinal neurons \cite{ming1997} as well as rat spinal neurons \cite{bouzigues2007} were also in the range of $\sim 20^o$. We consider this to be a strong, but qualitative validation of our model of MT polarization. 

Regarding the spatial sensitivity of the system, the model predicts that the attractive or repulsive cue needs to be spread across minimally $30^o$ angle of the growth cone. Most experiments of GC turning involve a pipette assay where the chemical that provides the directional cue is applied extracellularly centered $45^o$ to the GC and angular spread appears quite large ($\sim 90^o$) and is difficult to control in such {\it vitro} assays \cite{pujic2009}. However bead based assays \cite{moore2009} suggest an alternative method for estimating the role of spatial extent of the cue. The larger question of the {\it in vivo} scenario might also further complicate matters with multiple cues simultaneously stimulating a single GC. We cannot resolve this in our simulation. Our model does however predict a lower limit to to the MT polarization system in the neuronal GCs, which can be experimentally tested. The guidance of neurons was shown to increase linearly with signal to noise ratio (SNR) both in theory and experiment with rat dorsal root ganglion (DRG) explants \cite{mortimer2009}.  Our results are qualitatively comparable with previous data of increased MT polarization with increasing receptor concentration and broadening of the receptor signal. However a more detailed receptor-ligand dynamics model will be required for a direct comparison between the two models.

We further resolve our spatial MT polarization model of GCs by including a stathmin-like regulator of MT dynamics and its cyclic phosphorylation-dephosphorylation, since stathmin has been shown to modulate MTs and affect neurite outgrowth in rat hippocampal neurons \citep{morii2006a}. The mapping from phospho-stathmin concentration to MT dynamics parameters however has been measured only in Xenopus oocytes \cite{stathmin_cat} and we use the empirical fit to this data as a start point. This biologically realistic scenario is used to test the effect of different network topologies such as a feedback from MT tips to the reaction network and saturated reaction kinetics.We find these amplifications do not improve the angle of turning, as compared to the simple cyclic phosphorylation-dephosphorylation reaction system. The feedback from MTs intact appears in our system to amplify noise and cause `spurious' polarization even in the absence of a cue. On the other hand we do find that receptor activation strength proves to be a much better means of strengthening the polarization of the MT system. This result reveals an important design principle of such a system- namely amplification steps appear more effect at the receptor signaling level. This is consistent with recent findings in a simple model of MT feedback based GABA receptor polarization that amplifies receptor based signaling \cite{bouzigues2010}. Our study goes beyond this to analyze the explicit regulation of MT dynamics by both chemical and physical mechanisms. 

In conclusion, the model presented here predicts various mechanical and chemical properties of MT cytoskeleton regulation in neuronal GC turning. Validation with existing data demonstrates the utility of this model and points the way for additional experimental testing. The predicted physical limits to MT polarization in GCs is important, since it reveals potential design principles of neuronal GC turning. Additionally recent developments in microfluidics and micro patterning techniques \cite{wang2008, rosoff2004} suggest means of experimental testing of these predictions. The recent  {\it in vivo} evidence from neuronal regeneration by MT stabilization \cite{hellal2011} shows the growing need to understand the role of microtubule polarization at a system level in neurons. We believe this study is a step in this direction.

\section*{Acknowledgments}
The work was supported by core funding by IISER Pune, India. We thanks Francois Nedelec for discussions regarding cytoskeletal simulations.

\newpage
\singlespacing 
\bibliography{growthconeLit}

\clearpage
\newpage
\section*{Tables}

\begin{table}[ht]
\begin{tabular}{|c|c|c|c|}
  \hline
  {\bf Location of MT plus-end} & {\bf P-domain}     & {\bf T- and C-domain} & \bf Reference \\
  \hline
  $v_g$  $(\mu m-s^{-1})$ & $11.5 \cdot 10^{-2}$  & $6.4 \cdot 10^{-2}$  & \cite{schaefer2002}\\
   \hline
  $v_s$  $(\mu m-s^{-1})$ &  $20    \cdot 10^{-2}$  & $5.1 \cdot 10^{-2}$  &  "\\
   \hline
  $f_{cat}$ $(s^{-1})$         &  $1.7   \cdot 10^{-2}$  & $0.5 \cdot 10^{-2}$  &  "\\
   \hline
  $f_{res}$ $(s^{-1})$         &  $ 2.3  \cdot 10^{-2}$  & $9.5 \cdot 10^{-2}$  &  "\\
  \hline
\end{tabular}
\begin{flushleft}Parameters of Dynamic Instability
\end{flushleft}
\caption{\bf{The dynamic instability parameter values were chosen for the P- and C-domains and T-zone based on published values from {\it Aplysia} neuronal GCs.} }
\label{tab:mtdynpar}
\end{table}

\begin{table}[ht]
\begin{tabular}{|p{5cm}|p{3cm}|p{3cm}|p{2cm}|}
  \hline
{\bf Parameter} & {\bf P-domain }& {\bf T-zone } & {\bf Reference}\\ 
  \hline
   No. of motors &  30,000 & 1,000 & optimized \\
    \hline
    Speed $v_{mot}$ ($\mu m/min$) & 4.2 & 0 & \cite{schaefer2002} \\
  \hline
     Binding rate $r_{attach}$ $(s^{-1})$  & 5  & 5  &  optimized \\
  \hline
     Unbinding rate $r_{detach}^{'}$ $(s^{-1})$  & 1  &  1 &  optimized \\
  \hline
   Maximum stalling force $f_{stall}$ (pN) (Kinesin and actin-retrograde flow) & 15  & 15 & \cite{Kinesin_GC_turning, Retrograde_Actin_Flow_Model, coppin1997} \\
  \hline
     Attach distance $(\mu m)$  & 0.1 & 0.1 & optimized \\
  \hline
       Motor stiffness $k$ $(pN/\mu m$  & 100 & 100 & \cite{howard2001mechanics}, \cite{coppin1997}  \\
  \hline
\end{tabular}
\begin{flushleft} Motor parameters
\end{flushleft}
\caption{\bf{The parameters of surface-immobilized motors were chosen for the P-domain and T-zone to fit the retrograde flow rates and effective translocation speeds of MTs reported \cite{schaefer2002}.}}
\label{tab:motorparam}
\end{table}

\clearpage
\section*{Figure Legends}

\subsubsection*{Figure~\ref{model-schematic}.}
{\bf Geometry of growth cone MT-motor system} (A) A semi-circular geometry represents the growth cone, which is divided into three concentric regions: P-domain, T-zone and C-domain. The C-domain is connected to an axon shaft. A MT stabilization zone in the P-domain promotes MT polymerization. Immobilized plus-end directed motors are arranged in sixty-one symmetric radial arrays spaced at intervals of $3^o$ in the P-domain. Non-motile motors similarly immobilized are localized in the T-zone. The reaction-diffusion system is bounded in the light and dark regions (a subset of simulation space). An effective force field in the GC-neck and axon shaft represents a compressive force (white arrows). The spatially distributed dynamic instability parameters $v_g$, $v_s$, $f_{cat}$ and $f_{res}$ for the plus- and minus-ends of are plotted along the GC-axon axis. (B) The geometry of the modeled MT-motor interactions in the T-zone with non-motile complexes and in the P-domain with plus-end directed motors is schematically represented.
 
 \subsubsection*{Figure~\ref{fig:typical-simulation}.}
 {\bf MT dynamics in an unbiased growth cone} (A) The time-series of MTs entering the P-domain (the region with higher catastrophe frequencies) in simulation snapshots at 0, 600 and 12000 s is represented in 2D plots and (B) as a time course of MTs entering the P-domain and achieving a fluctuating steady-state. (C) At steady-state the number of MTs in the P-domain in simulations with 30 or 60 initialized MTs was compared to images from experiments (n=2) \cite{schaefer2002, bundling_forces}. (D) The geometry of the growth cone angular sectors is schematically represented. (E) This is used to calculate the time-dependent mean angle of MTs $(\theta_{av}(t))$and its standard deviation $(\sigma_c(t))$. (F) The simulated mean angle and spread at steady-state ($t >400s$) was compared to experimental distributions (n=2) \cite{schaefer2002}.
 
 \subsubsection*{Figure~\ref{fig:retrograde-rates}.}
 {\bf Effects of perturbing the number of MTs and retrograde translocation rates} MT densities in the peripheral 25$\%$ of the growth cone are plotted against time for initial nucleated MTs numbering (A) 30, (B) 60 and (C) 90. (D) The number of MTs entering the P-domain per second show a $50\%$ change when standard retrograde translocation speeds (4.2 $\mu m/min$) are changed by $50\%$. This effect is visible in the most-distal quarter of the P-domain and not as pronounced when larger elements $(50-75\%)$ of the P-domain are sampled. (E) The absolute number of MTs in the P-domain are less dramatically affected by changes in retrograde translocation speeds. Error bars are standard deviation (Iterations n=10). Stars indicate statistically significant differences from the control case. 

 \subsubsection*{Figure~\ref{fig:polarization_parameters}.}
{\bf MT polarization in a biased growth cone} (A) The time-dependent mean angle $(\theta_{av}(t))$ of MTs shows a change in mean direction when a directional bias is introduced in the left half of the GC at 800s by decreasing $f_{cat}$ ten-fold. (B) The angular frequency distribution of the mean number of MTs at steady state shows a clear polarization in the biased case. The sectors are $15^{o}$ apart while the radial distance from the origin indicates the number of MTs (Iterations n=10). (C) The mean change in angle of MTs after the bias ($\Delta\langle\theta_{av}\rangle$) is the greatest for a a ten-fold increase in the $f_{cat}$ values in one half of the GC, but the change is also statistically significant (star) for an increase in $f_{res}$ and decrease in $f_{cat}$. (D) Fold changes in $f_{cat}$ and $f_{res}$ biases (Iterations n=10) show a statistically significant changes (star) for two-fold or greater changes in both parameters. Error bars indicate standard errors (n=10).

 \subsubsection*{Figure~\ref{fig:polarization-sens}.}
 {\bf Spatial and temporal sensitivity} Decreasing angular sectors of the GC were biased by changing values ten-fold of (A) $f_{res}$ (increased) and (B) $f_{cat}$ (decreased). The sectors were centered around $120^{\circ}$ (striped) with respect to the GC semi-circle. Less broad biases in $f_{cat}$ resulted in a steep reduction in MT polarization, while it remained fairly similar for $f_{res}$ biases as narrow as $10^o$. Error bars indicate standard errors (n=10). (C) The time taken for the half-maximal change in mean MT angle after a bias is introduced ($\Delta K_t$) was estimated by fitting the time-course of mean MT angle ($\theta_{av}(t)$). (D) The most rapid change in MT mean angle was produced by a ten-fold reduction in $f_{cat}$, while a similar reduction in $f_{res}$ produced almost no change.

 \subsubsection*{Figure~\ref{fig:rd-optimization}.}
{\bf Spatial reaction network} (A) A schematic view of the phosphorylation of the regulatory protein $S$ by a membrane receptor signal $R$ converts it to $S^*$. S itself increases the frequency of catastrophe. The dotted lines indicate the additional feedback of MTs that promotes the $S$ to $S^*$ transition, further stabilizing MTs. The intensity of brightness (dark: low, bright: high) in the GC plots the spatial extent of $S$. (B) Lines indicating the (C) radial and (D) vertical (GC to axon) axes along which concentrations were plotted. Solving the simple cyclic reaction network, the initial and steady-state concentration profiles of $S$ and $S^*$ show a gradient from the GC periphery inwards along both (C) radial and (D) vertical axes. The corresponding $f_{cat}$ gradient profiles result from the concentration gradient of the network.

 \subsubsection*{Figure~\ref{fig:feedback}.}
 {\bf The amplification of the signaling system} (A) The polarization of MTs in response to zero-order ($K_{M1}/S_{tot}=0.1$), intermediate ($K_{M1}/S_{tot}=1$) and first-order ($K_{M1}/S_{tot}=0.1$) reaction kinetics are compared to increasing values of the receptor signal (n=30). (B) The schematic polar frequency distribution of the number of MTs  shows the greatest polarization in the MTs in response to a first-order reaction kinetic with the highest receptor signal. (C) Spatial plots of $S$ distribution in the GC are compared for two networks- the simple cyclic reaction without feedback ($R=200 \mu M$, $K_{M1}/S_{tot}=2$, $n_1=1$) and the growing MT-tip driven positive feedback loop (with additional parameters $w_1=200$, $K_{M2}/S_{tot}=0.1$, $n_2=10$). (D) The polar plots at steady state of number of MTs display a highly polarized distribution for the simple cyclic reaction without feedback as compared to the strong feedback network.

\clearpage
\newpage
\section*{Figures}


\renewcommand{\thesubfigure}{\Alph{subfigure}}

\begin{figure}[ht]
\begin{center}
\leavevmode
	\subfigure{{\bf{A}}	 \label{fig:dynfield} }	
	\subfigure{{\bf{B}}	\label{fig:mt-actin} }	
	\includegraphics[width=0.5\textwidth]{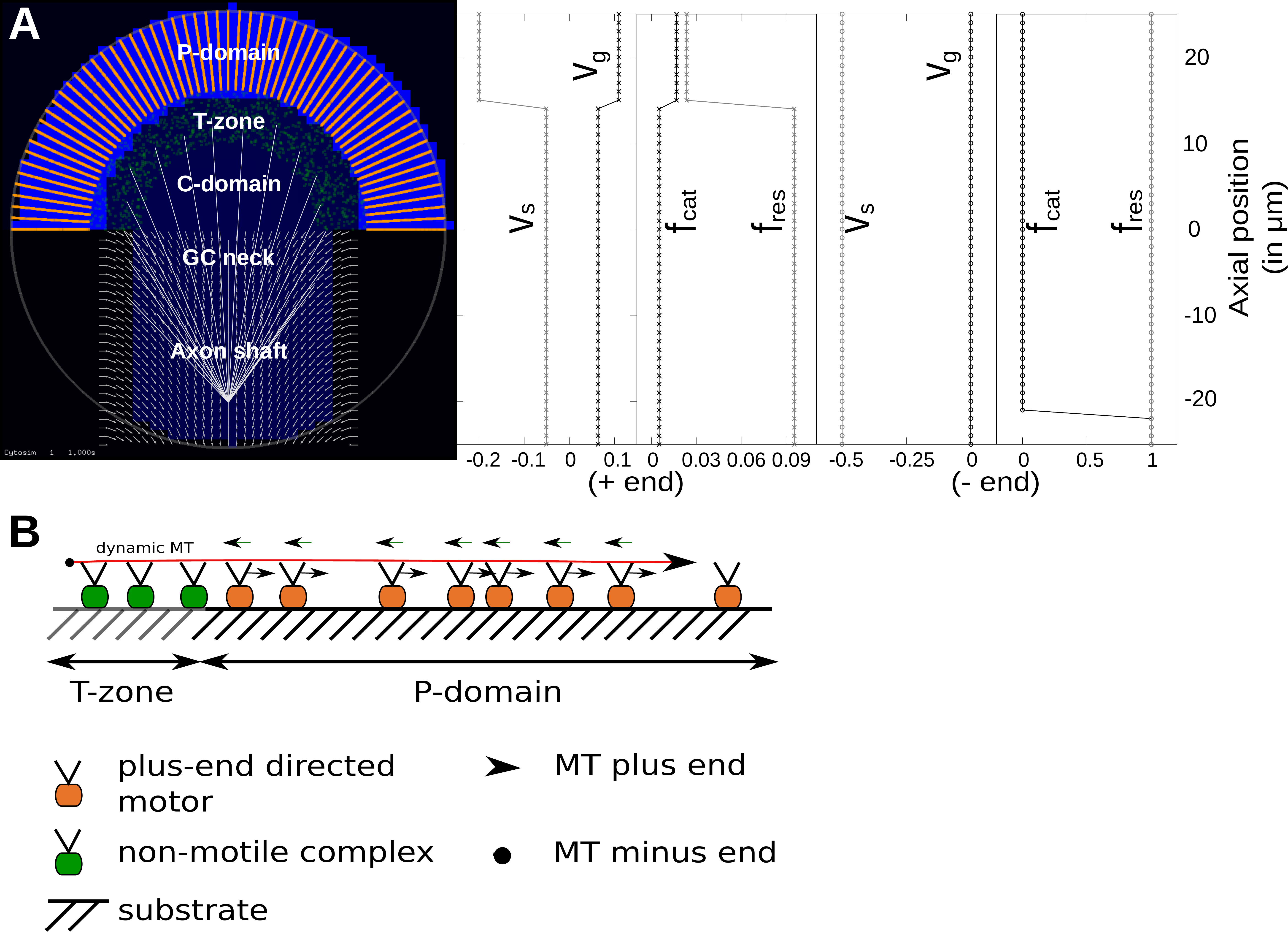}
\end{center}
\caption{}
\label{model-schematic}    
\end{figure}

\begin{figure}[ht]
\centering    
	\subfigure{{\bf{A}}	\label{fig:gcdyn1} }	  
	\subfigure{{\bf{B}} 	\label{fig:gcdyn2} 	}  
	\subfigure{{\bf{C}}		 \label{fig:gcdyn3} }
	\subfigure{{\bf{D}}   \label{fig:gcdyn4} }	     
	\subfigure{{\bf{E}}	\label{fig:gcdyn5} 	}     
	\subfigure{{\bf{F}}		\label{fig:gcdyn6} } 
	\includegraphics[width=0.5\textwidth]{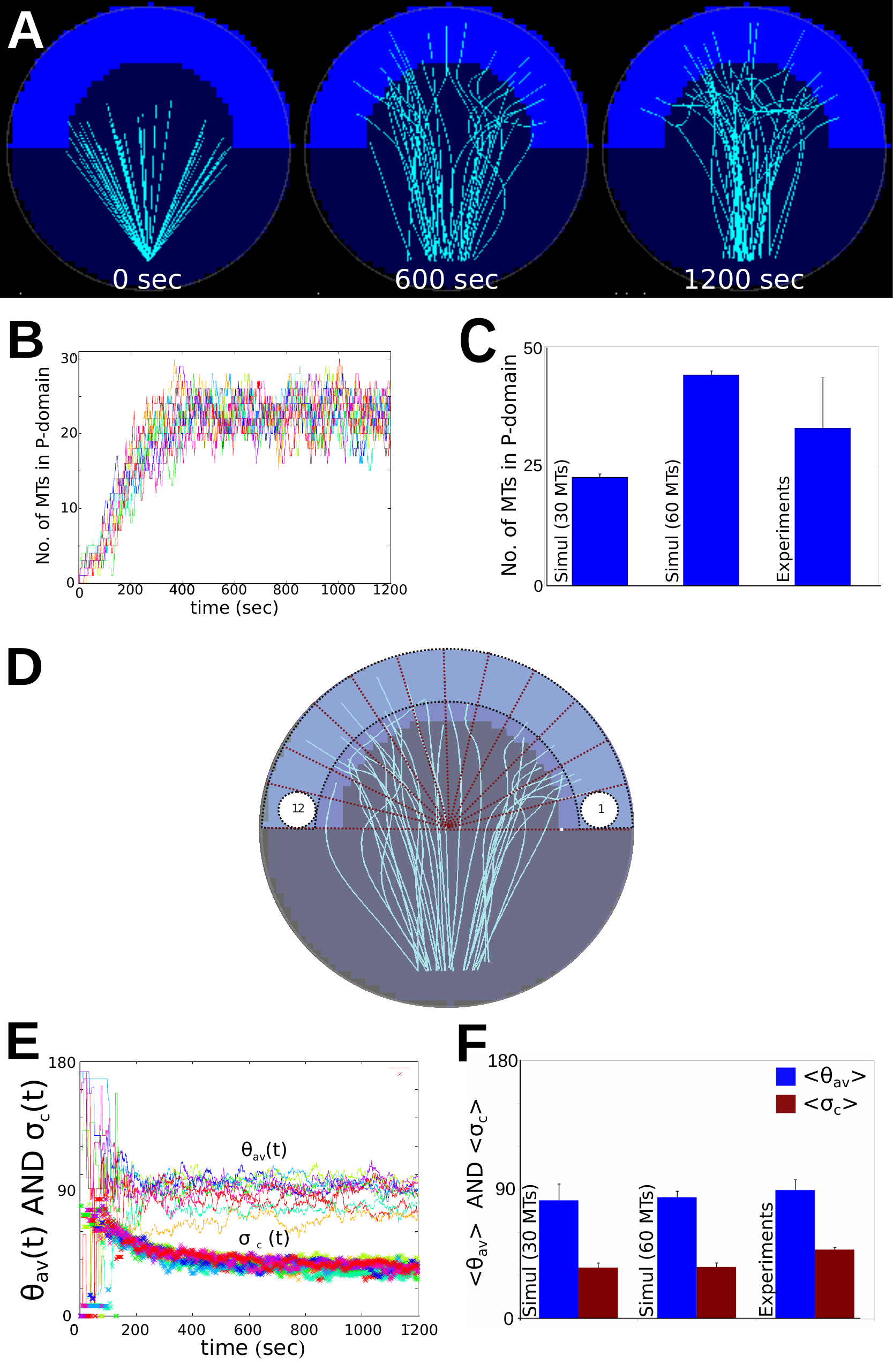}
\caption{}
\label{fig:typical-simulation}    
\end{figure}

\begin{figure}[ht]
\begin{center}
\leavevmode
	\subfigure{{\bf{A}}		\label{fig:nmt30} }      
	\subfigure{{\bf{B}}		\label{fig:nmt60} }     
	\subfigure{{\bf{C}}		\label{fig:nmt90} }       
	\subfigure{{\bf{D}}		\label{fig:flux-retrograde1} }	
	\subfigure{{\bf{E}}		\label{fig:flux-retrograde2} }	
	\includegraphics[width=0.5\textwidth]{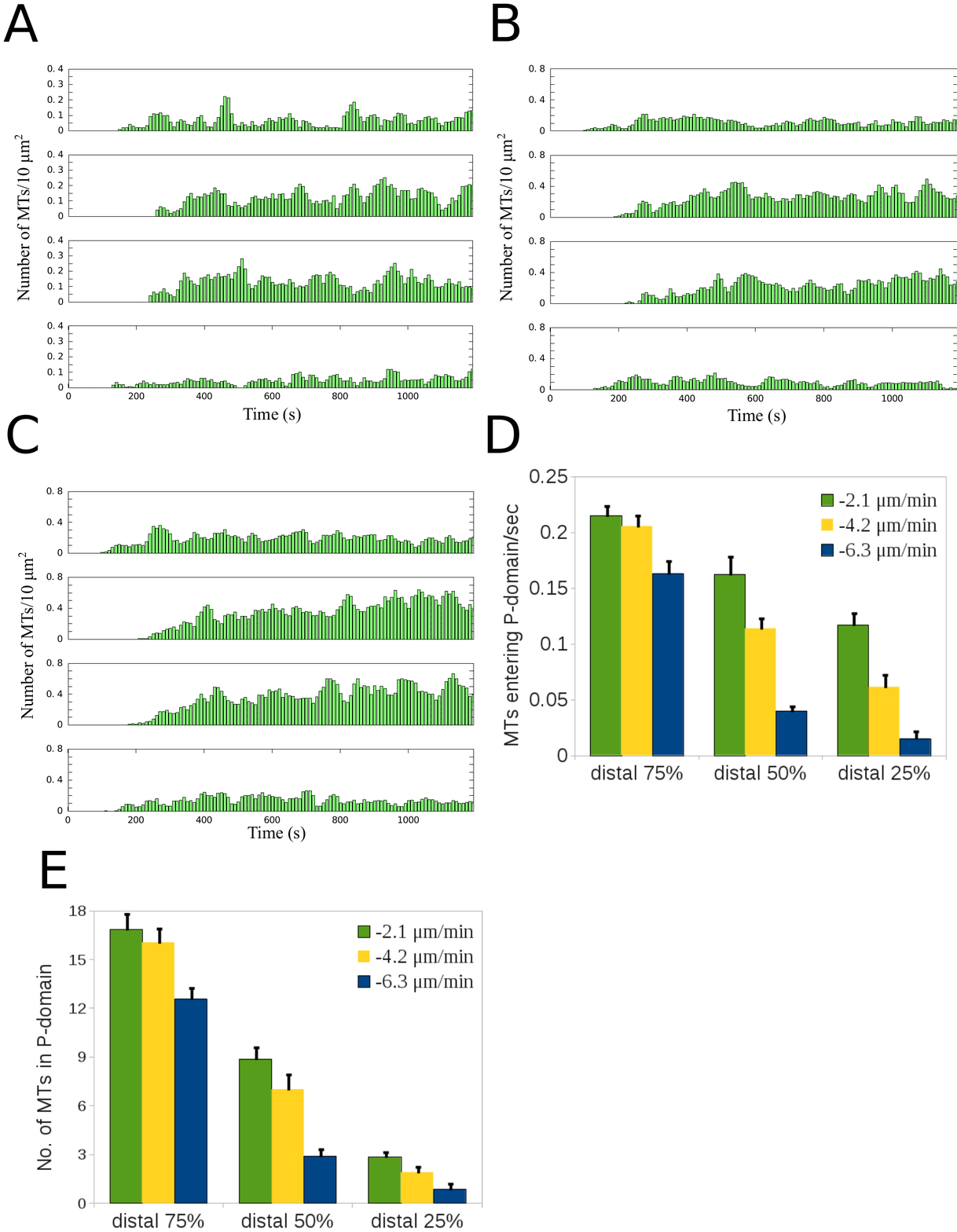}
\end{center}
\caption{}
\label{fig:retrograde-rates}
\end{figure}

\begin{figure}[ht]
\begin{center}
\leavevmode
	\subfigure	{{\bf{A}}	\label{fig:polarization-theta1} }	
	\subfigure	{{\bf{B}}	\label{fig:polarization-theta2} }	
	\subfigure	{{\bf{C}}	\label{fig:polarization-theta3} }	
	\subfigure	{{\bf{D}}	\label{fig:polarization-theta4} }	
	\includegraphics[width=0.5\textwidth]{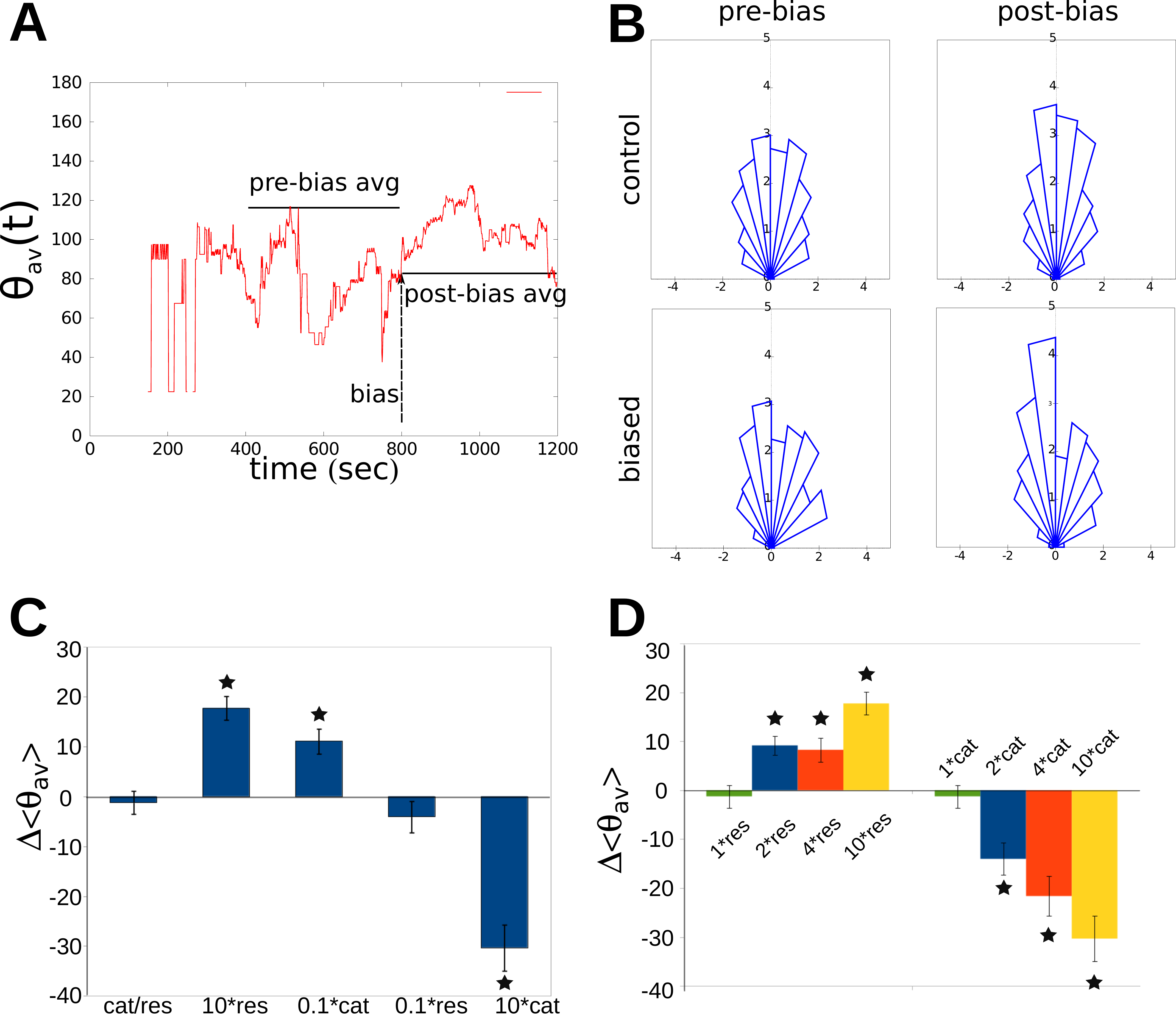}
\end{center}
\caption{}
\label{fig:polarization_parameters}
\end{figure}

\begin{figure}[ht]
\begin{center}
\leavevmode
	\subfigure{{\bf{A}}		\label{fig:angleExtent2} }	
	\subfigure{{\bf{B}}		\label{fig:angleExtent1} }	
	\subfigure	{{\bf{C}}	\label{fig:angleExtent3} 	}	
	\subfigure	{{\bf{D}}	\label{fig:angleExtent4} 	}	
	\includegraphics[width=0.5\textwidth]{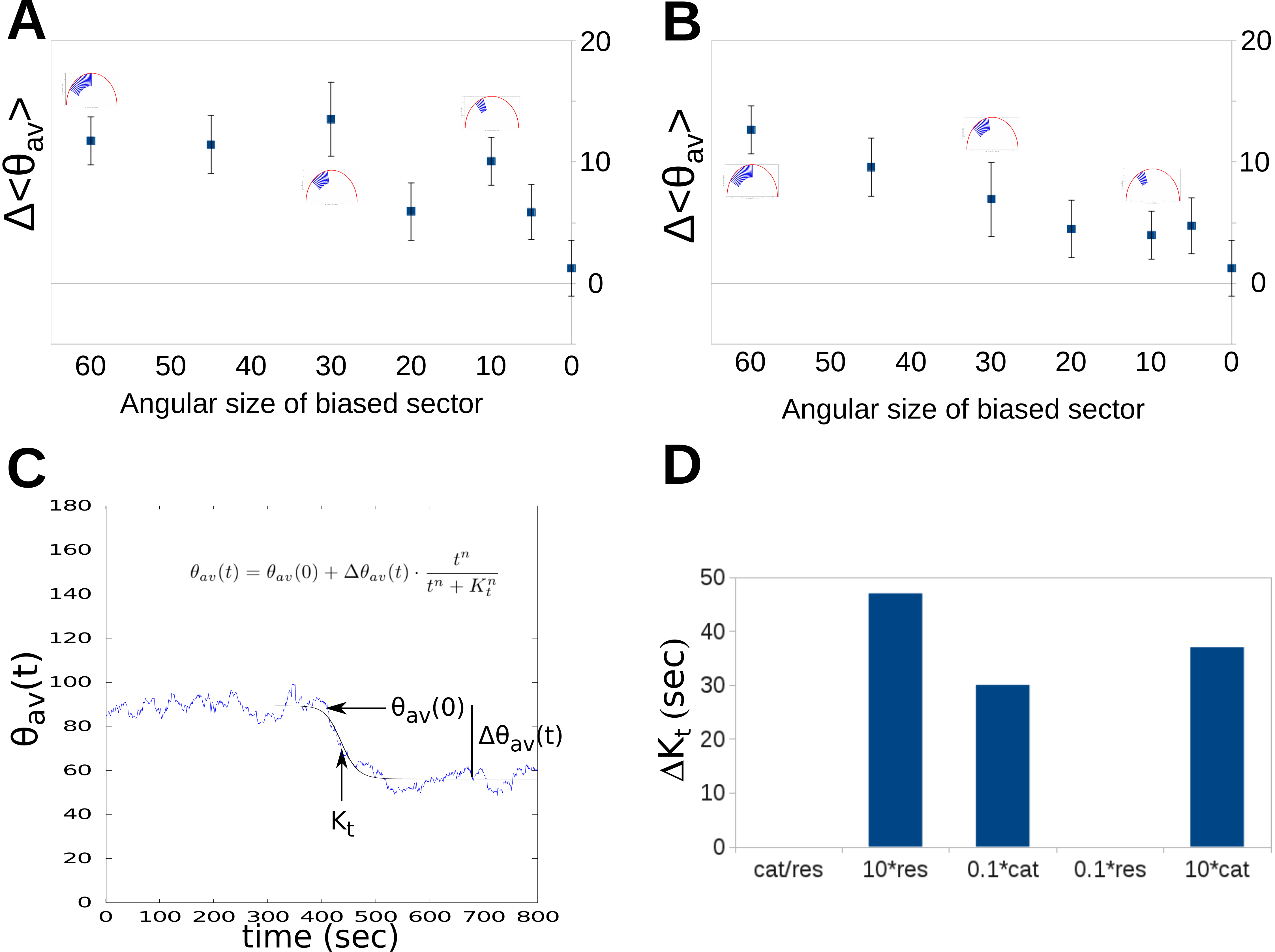}
\end{center}
\caption{ }
\label{fig:polarization-sens}
\end{figure}

%

\begin{figure}[ht]
\begin{center}
	\subfigure{{\bf{A}}		\label{fig:reac-scheme1} 	}	
	\subfigure{{\bf{B}}		\label{fig:reac-scheme2} 	}	
	\subfigure{{\bf{C}}		\label{fig:reac-scheme3} 	}	
	\subfigure	{{\bf{D}}	          \label{fig:reac-scheme4}	}	
	\includegraphics[width=0.5\textwidth]{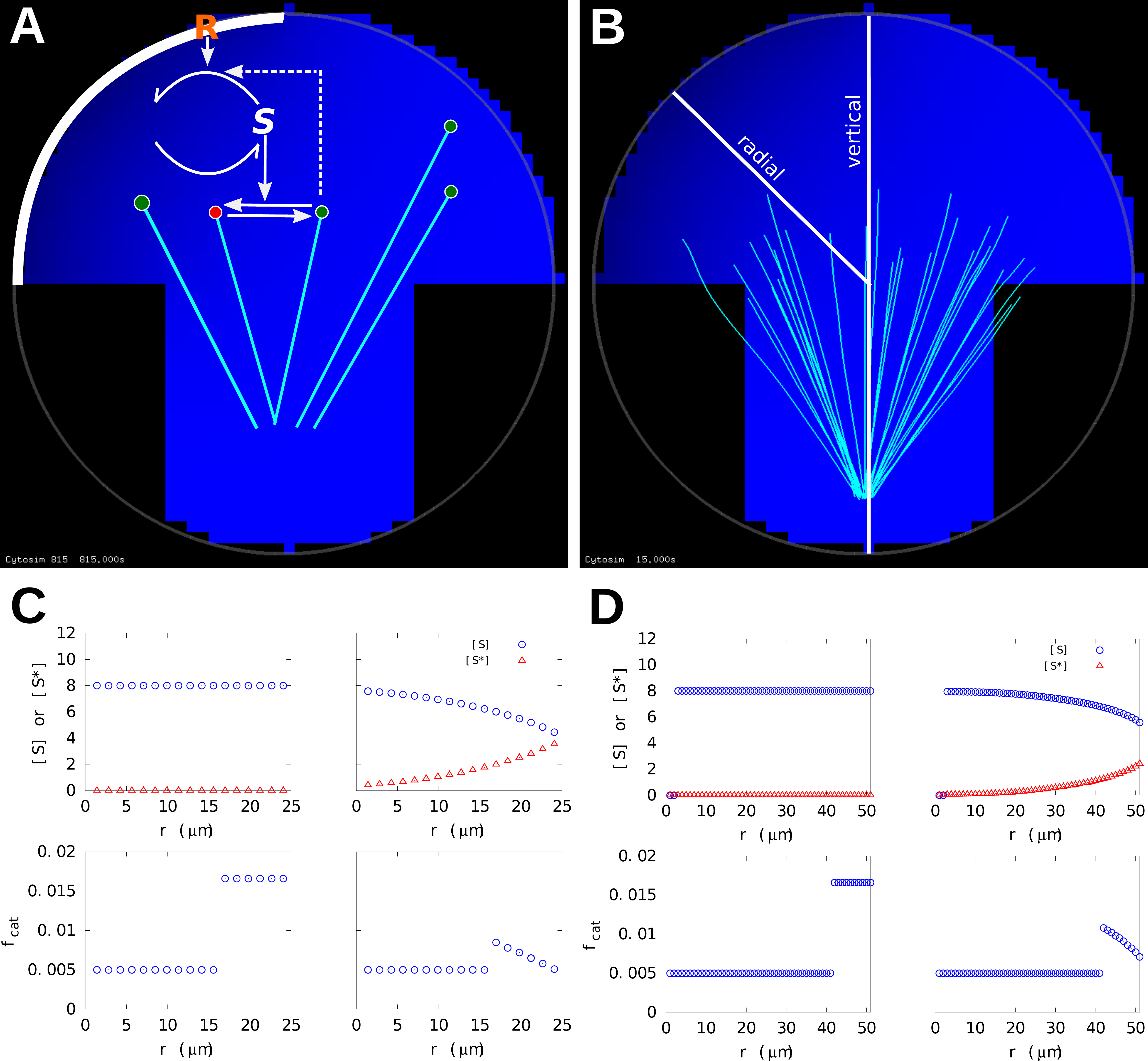}
\end{center}
\caption{}
\label{fig:rd-optimization}
\end{figure}


\begin{figure}[ht]
\begin{center}
\leavevmode
	\subfigure{{\bf{A}}		\label{fig:Rsig-var1} 	}	
	\subfigure{{\bf{B}}		\label{fig:Rsig-var2} 	 }        
	\subfigure{{\bf{C}}		\label{fig:feedback1}	 }	
	\subfigure{{\bf{D}}		\label{fig:feedback2}	 }	
	\includegraphics[width=0.5\textwidth]{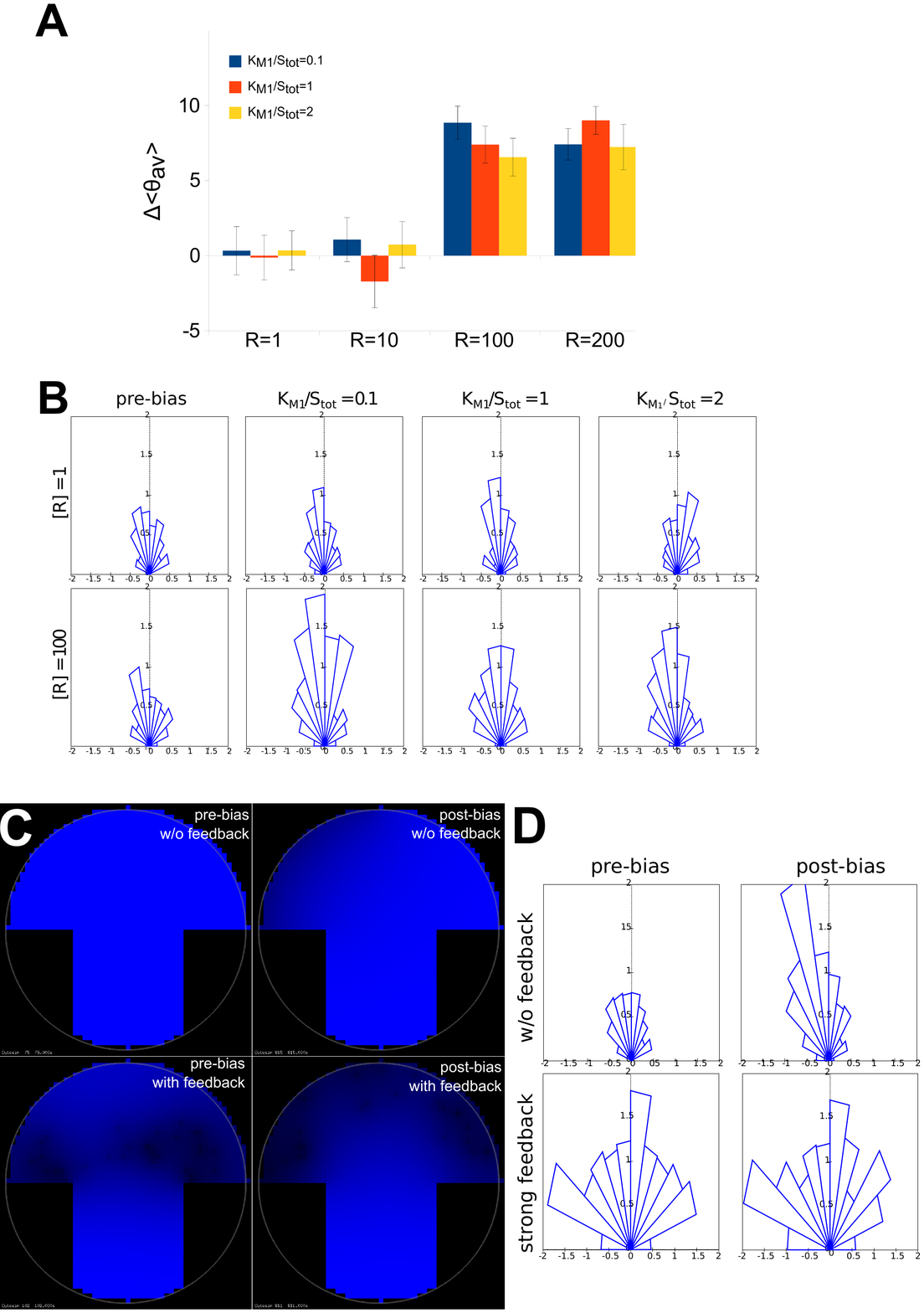}
\end{center}
\caption{}
\label{fig:feedback}
\end{figure}


\clearpage
\newpage4
\section*{Supporting Figures}
\setcounter{figure}{0}
\makeatletter 
\renewcommand{\thefigure}{S\@arabic\c@figure}

\begin{figure}[ht]
\begin{center}
\includegraphics[width=0.75\textwidth]{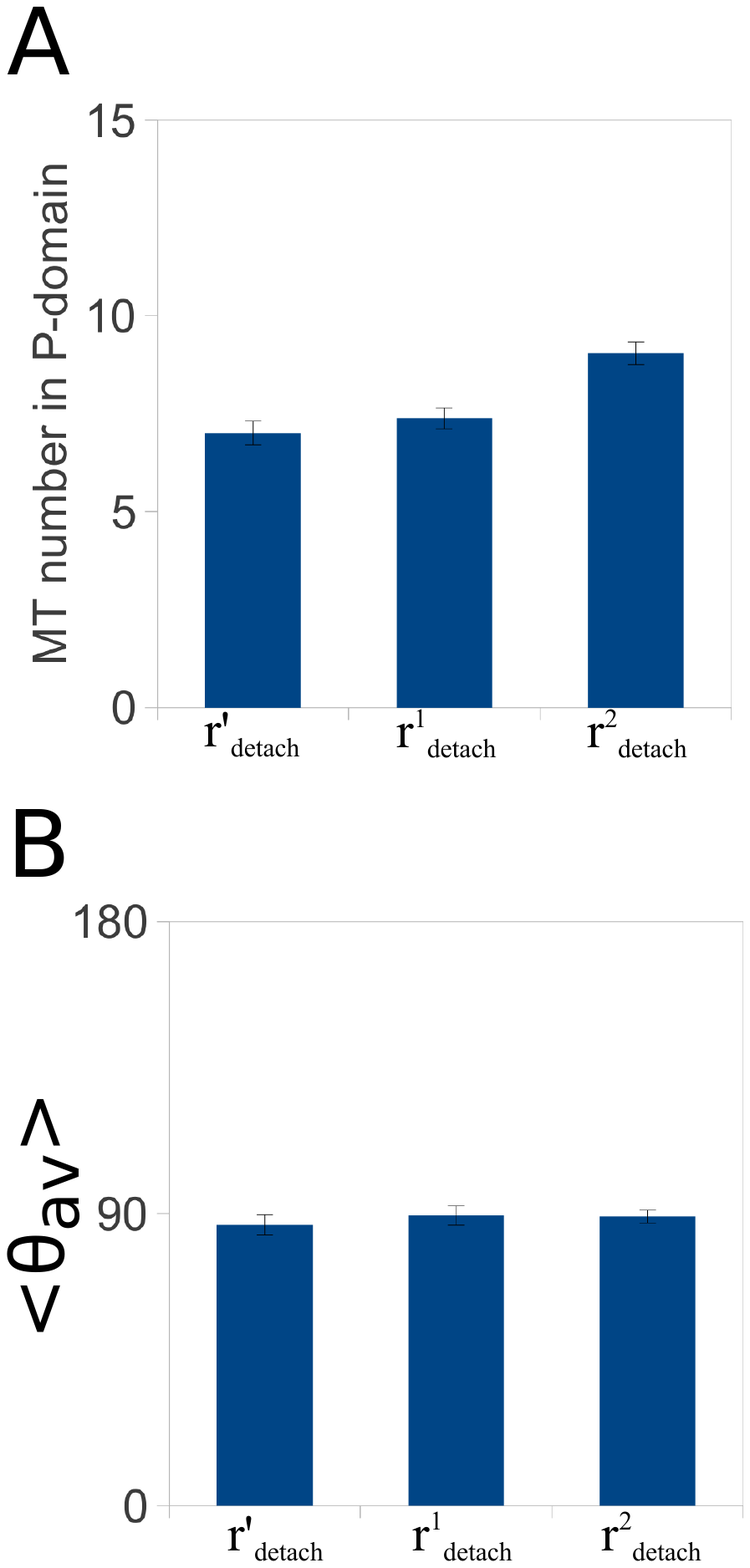}
\end{center}
\caption{}
\label{sup:detachmodel}
\end{figure}

\begin{figure}[ht]
\begin{center}
\includegraphics[width=0.75\textwidth]{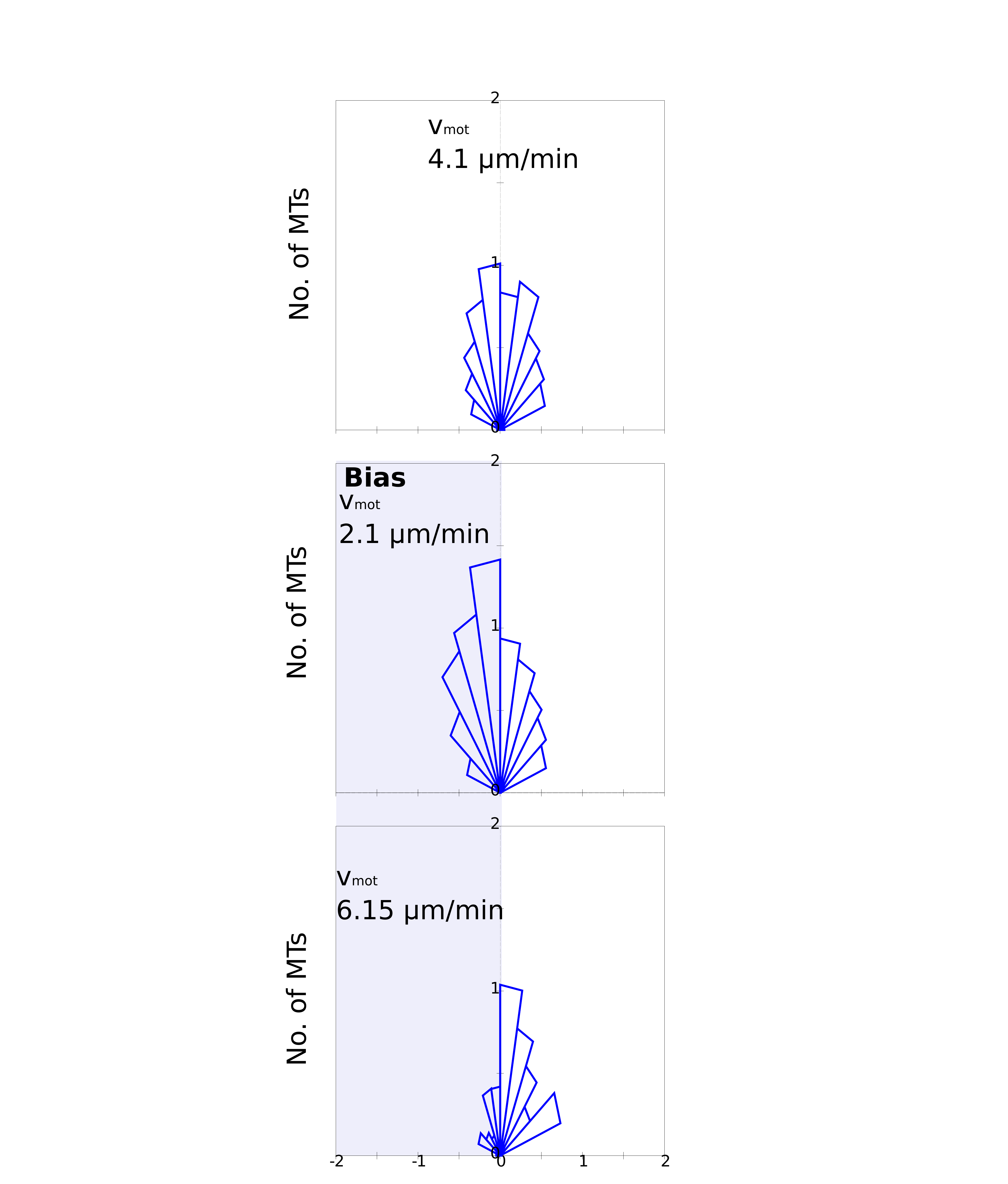}
\end{center}
\caption{}
\label{sup:retrovel}
\end{figure}

\begin{figure}[ht]
\begin{center}
\includegraphics[width=0.75\textwidth]{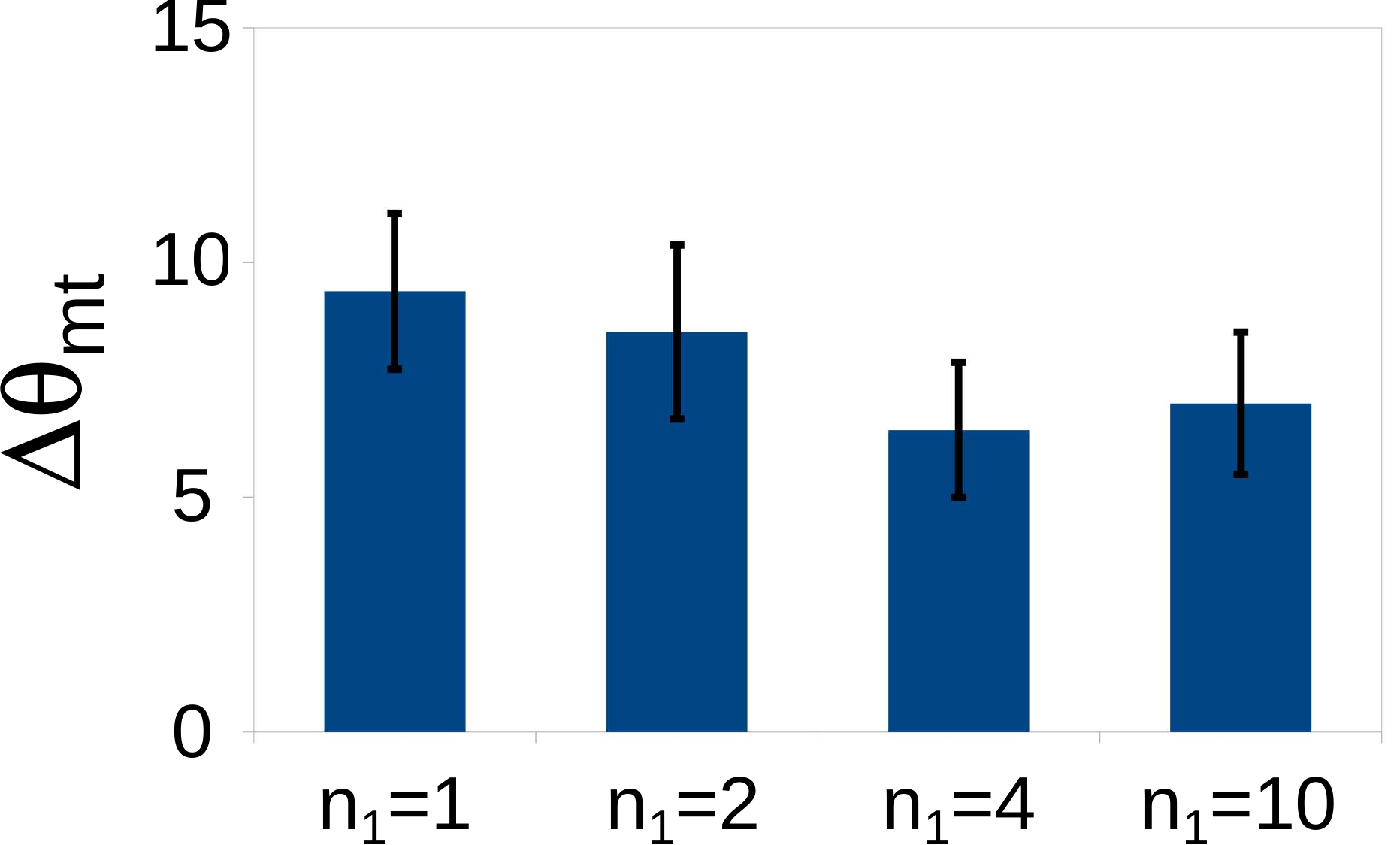}
\end{center}
\caption{}
\label{sup:hillcoef}
\end{figure}

\begin{figure}[ht]
\begin{center}
\includegraphics[width=0.75\textwidth]{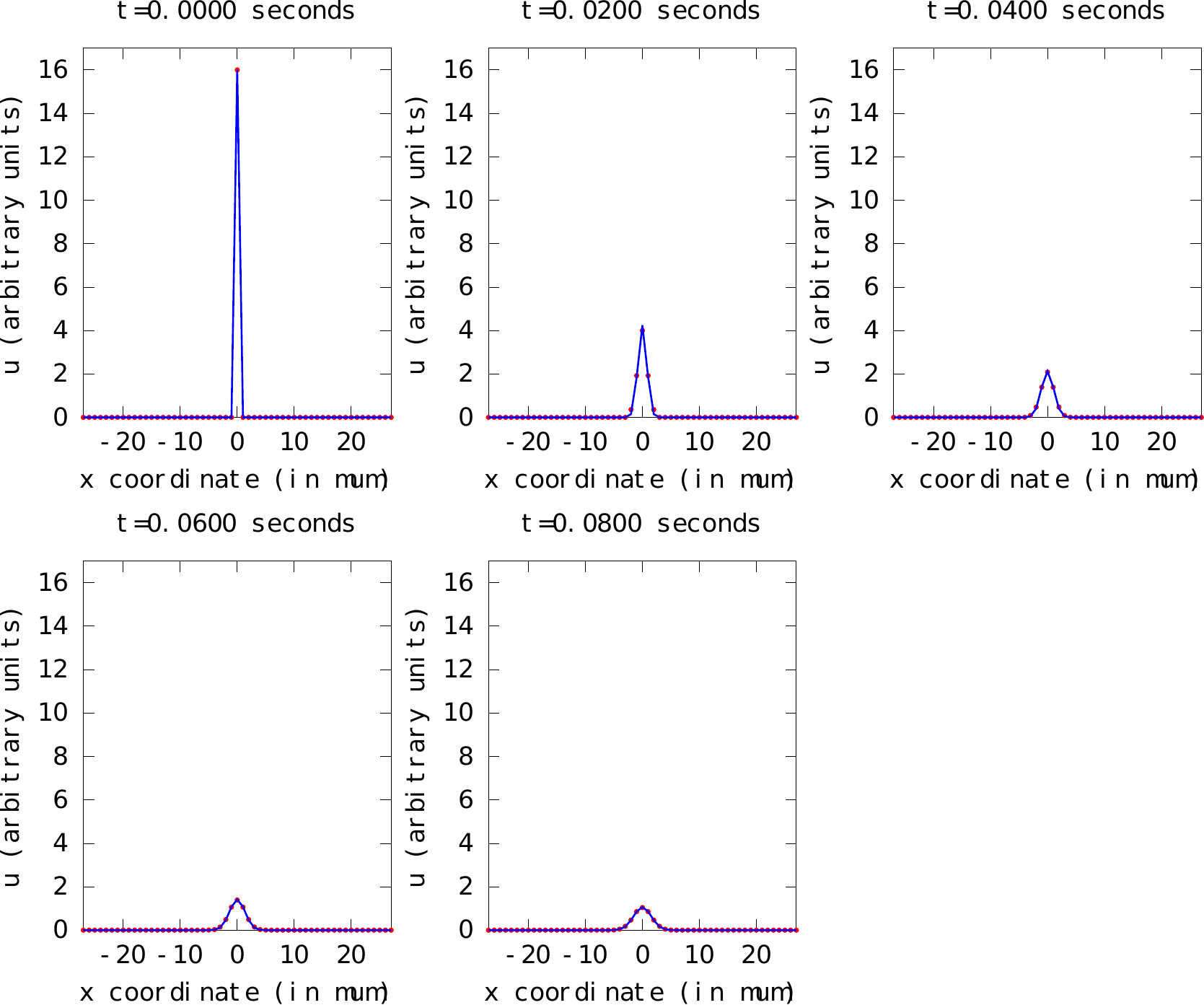}
\end{center}
\caption{}
\label{sup:rd-optimization}
\end{figure}


\clearpage
\newpage
\section*{Supporting Videos}
\setcounter{figure}{0}
\makeatletter 
\renewcommand{\thefigure}{S\@arabic\c@figure}

\begin{figure}[ht]
\begin{center}
\includegraphics[width=0.75\textwidth]{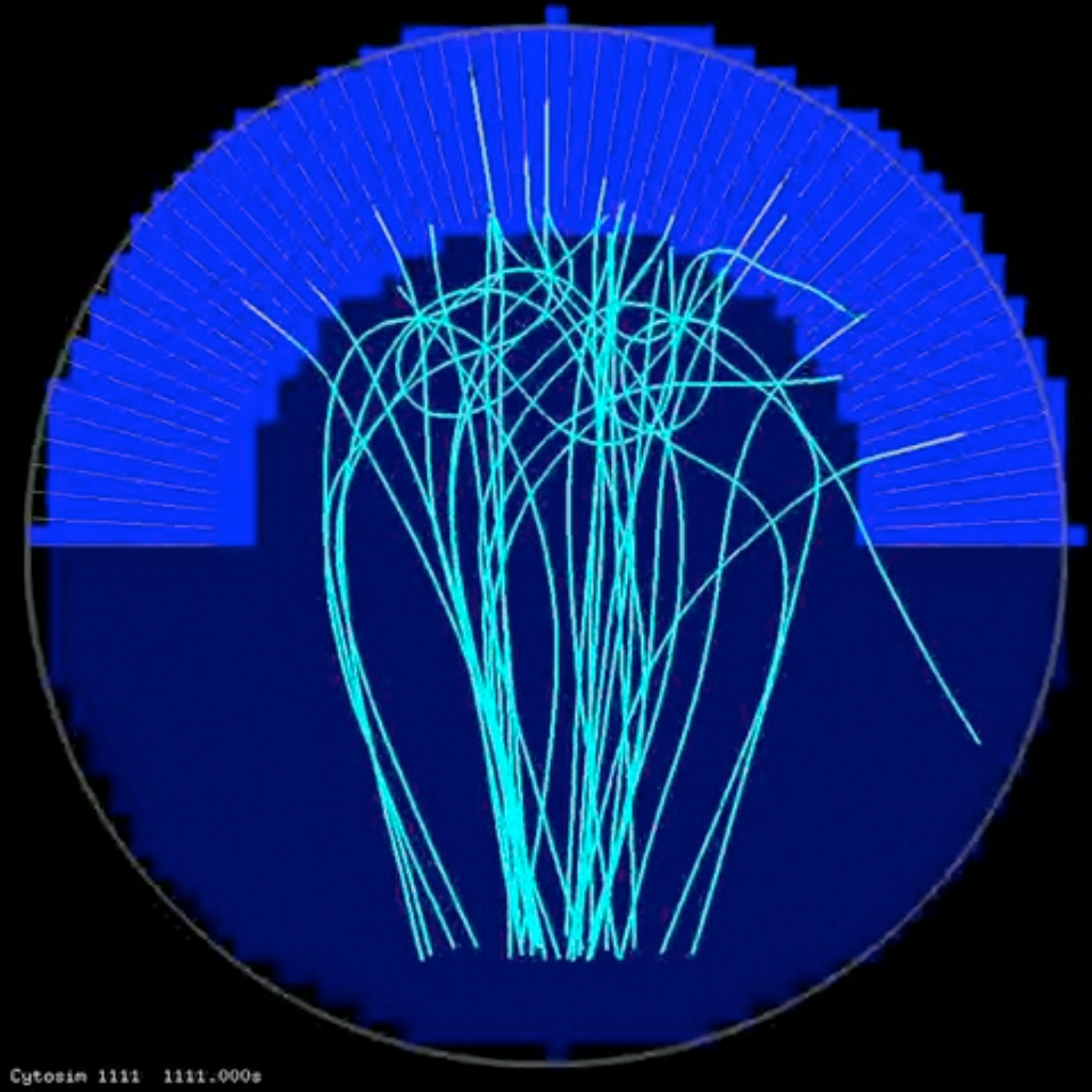}
\end{center}
\caption{}
\label{sv1}
\end{figure}

\begin{figure}[ht]
\begin{center}
\includegraphics[width=0.75\textwidth]{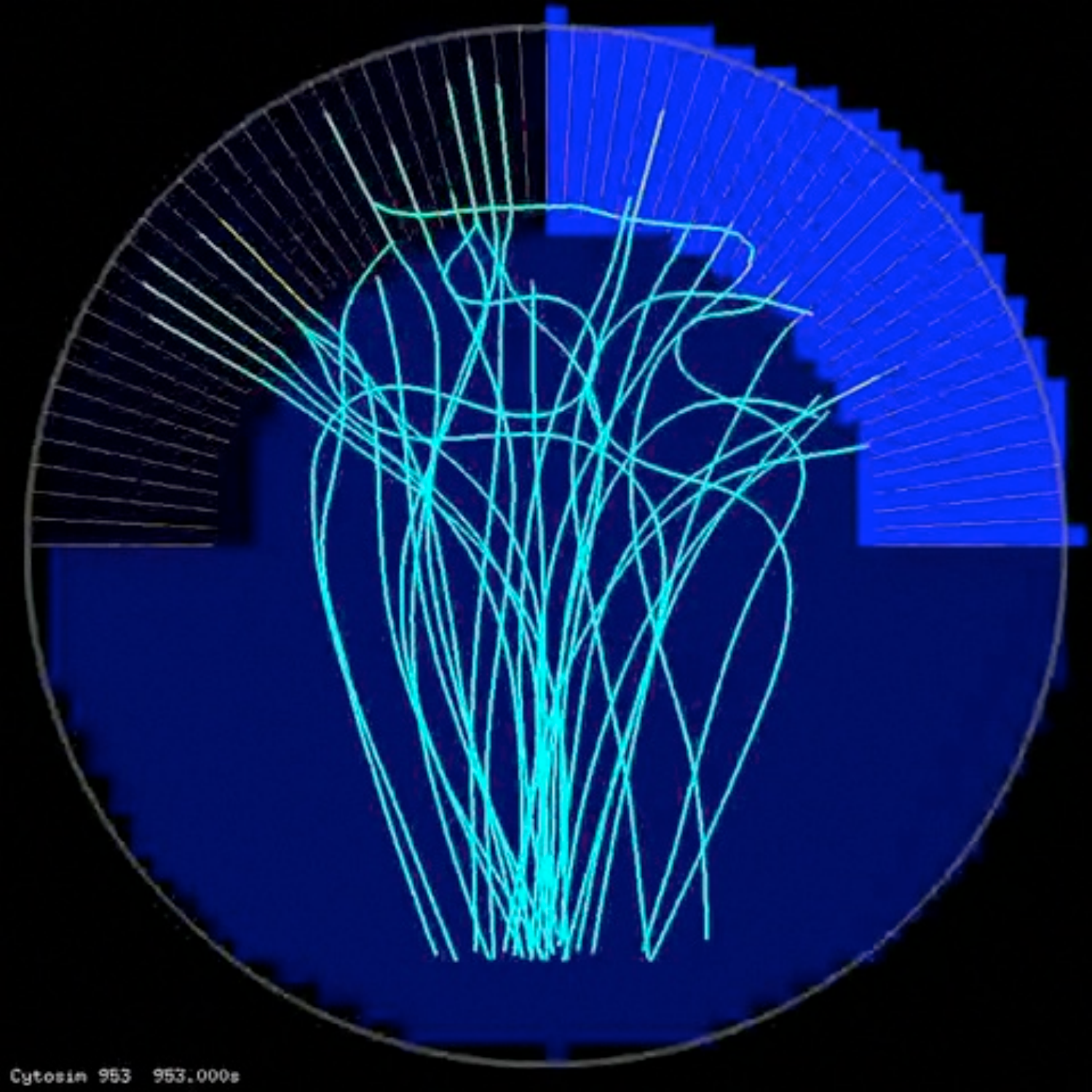}
\end{center}
\caption{}
\label{sv2}
\end{figure}

\begin{figure}[ht]
\begin{center}
\includegraphics[width=0.75\textwidth]{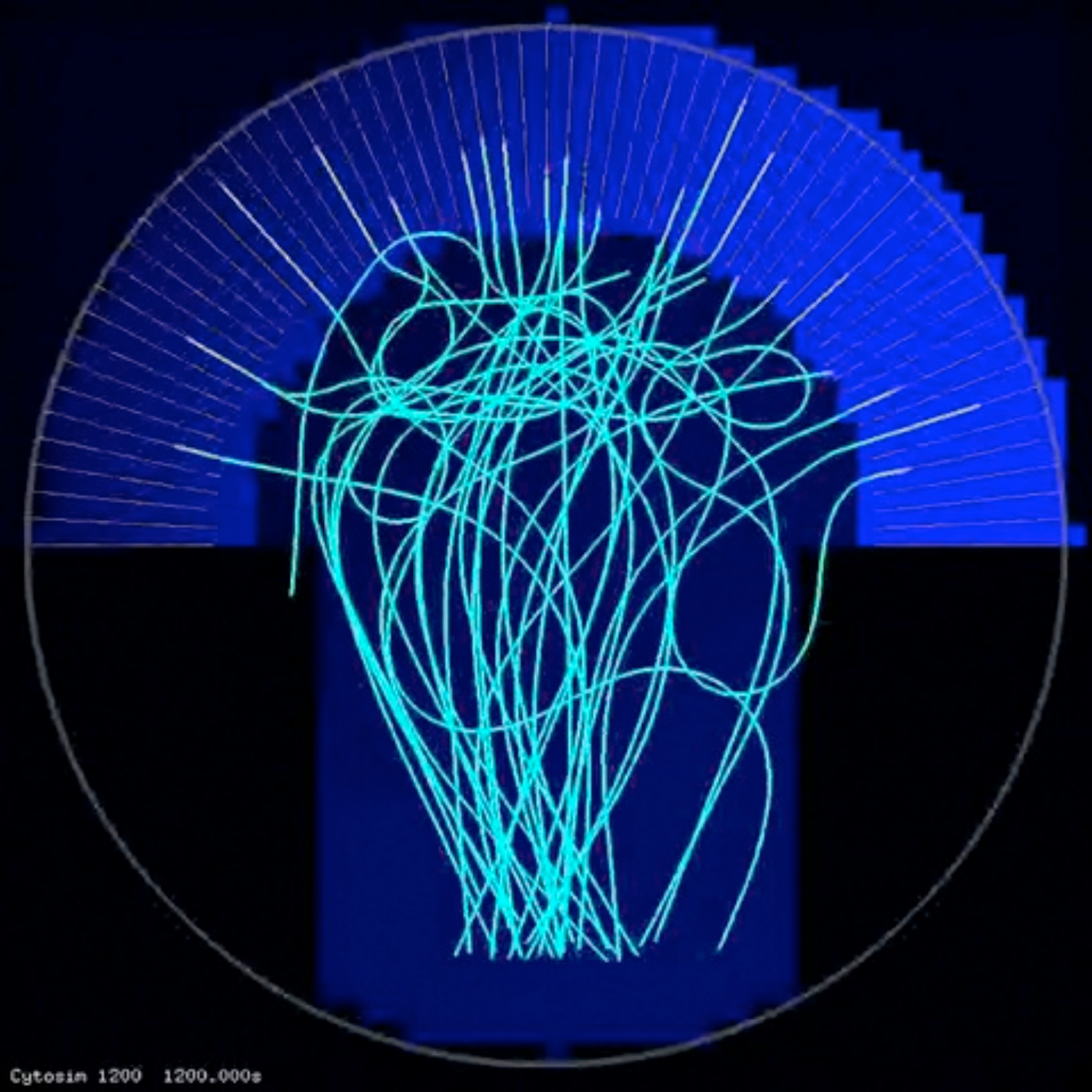}
\end{center}
\caption{}
\label{sv3}
\end{figure}


\clearpage
\newpage
\section*{Supplementary Tables}
\setcounter{table}{0}
\makeatletter 
\renewcommand{\thetable}{S\@arabic\c@table} 

\begin{table}[ht]
\begin{tabular}{|p{1cm}|p{3.5cm}|p{3.5cm}|p{3.5cm}|p{3.5cm}|p{3.5cm}|}
  \hline
  {\bf Parameter} & {\bf Description} &  {\bf Value} & {\bf Unit} & {\bf Reference}\\ 
  \hline
  {\bf $[S]_{tot}$ } &Total stathmin concentration &  8 & $\mu M $ &\cite{stathmin_cat}\\
  \hline
  {\bf $D_S$ } & Diffusion coefficient of $S$  &  15 &  $\mu m^2 s^{-1}$ &\cite{stathmin_grad}\\
    \hline
    {\bf $D_{S^*}$ } & Diffusion coefficient of $S^*$ &  15 &  $\mu m^2 s^{-1}$ &\cite{stathmin_grad}\\
  \hline
  {\bf $[R]$ } & Receptor concentration &  [1, 10, 100, 200] & $\mu M $ & -\\
  \hline
  {\bf $w_1$ } & Weight of positive feedback &  [1, 10, 100] &	$\mu M$ & -\\
  \hline
  {\bf $w_{2}$ } &  Weight of $f_{cat}$ dependence on [S] &  0.3144 & $ - $ &	\cite{stathmin_cat}\\
  \hline
  {\bf $k_{1}$ } & Forward rate constant &  0.1 & $\mu M^{-1}s^{-1} $ &	-\\
  \hline
  {\bf $k_{2}$ } & Backward rate constant &  0.1 & $s^{-1} $ &	-\\
  \hline
  {\bf $k_{3}$ } & Feedback rate constant  &  0.1 & $\mu M^{-1}s^{-1} $ &	-\\
  \hline
  {\bf $K_{M1}$ } & Michaelis-Menten constant ($K_M$) for the forward reaction &  [0.8, 8, 16] &	$\mu M $ & -\\
  \hline
  {\bf $K_{M2}$ } & Michaelis-Menten constant ($K_M$) for the feedback reaction &  8 & $\mu M $ &-\\
  \hline
  {\bf $n_{1}$ } & Hil Coefficient of the forward reaction &  [1, 2, 4, 10] & $-$ &   -\\
  \hline
  {\bf $n_{2}$ } &  Hill Coefficient of the feedback reaction &  [1, 10] &- & -\\
  \hline
\end{tabular}
\begin{flushleft}Reaction diffusion network parameters
\end{flushleft}
\caption{{\bf The parameters of the reaction diffusion scheme were chosen either based on previous experimental measurements or order of magnitude estimates. }}
\label{tab:rd-parameters}
\end{table}

\begin{table}[ht]
\begin{tabular}{|p{3.5cm}|p{3.5cm}|p{3.5cm}|p{3.5cm}|}
 \hline
  {\bf Region of P-domain considered (w.r.t periphery)} & {\bf Circumference ($\mu m$}) &  {\bf MTs crossing \underline{into the P-domain} per minute} & {\bf MTs entering the P-domain per 10 $\mu m-min^{-1}$}\\ 
  \hline
 100$\%$ (r=15 $\mu m$) & 47.1 & 14.4 & 3.06 \\
  \hline
  75$\%$ (r=17.5 $\mu m$) & 54.95 & 12 & 2.18 \\
  \hline
 50$\%$ (r=20 $\mu m$) & 62.8 & 6.6 & 1.05 \\
  \hline
   25$\%$ (r=22.5 $\mu m$) & 70.65 & 3.6  & 0.51 \\
  \hline

 \end{tabular}
\begin{flushleft}Flux of MTs entering the P-domain
\end{flushleft}
\caption{{\bf The parameters of the reaction diffusion scheme were chosen either based on previous experimental measurements or order of magnitude estimates. }}
\label{tab:mtflux}
\end{table}

\end{document}